\documentclass[a4paper,UKenglish,cleveref, autoref, thm-restate,]{lipics-v2021}
\title{Security Analysis of Bitcoin's V2 Transport Protocol: Exploiting Design
Implications for Sustained Eclipse and Downgrade Attacks}
\titlerunning{Security Analysis of Bitcoin's V2 Transport Protocol}

\author{Charmaine Ndolo}{Dresden University of
Technology}{charmaine.ndolo@tu-dresden.de}{https://orcid.org/0009-0003-6842-9444}{}
\author{Florian Tschorsch}{Dresden University of
Technology}{florian.tschorsch@tu-dresden.de}{https://orcid.org/0000-0001-6716-7225}{}
\authorrunning{C. Ndolo and F. Tschorsch}

\Copyright{Charmaine Ndolo and Florian Tschorsch}


\supplement{Source code used in this work is available as detailed in
\cref{sec:open_science}.}

\ArticleNo{1}

\usepackage{booktabs}
\usepackage{tabularx}
\usepackage[babel=true]{csquotes}
\usepackage{csvsimple}
\usepackage{etoolbox}
\usepackage{fontawesome5}
\usepackage[abbreviations]{glossaries-extra}
\glssetcategoryattribute{acronym}{indexonlyfirst}{true}
\usepackage{mathtools}
\usepackage{siunitx}
\DeclareSIUnit{\million}{M}
\usepackage[colorinlistoftodos, textsize=tiny,]{todonotes}
\usepackage{tabulary}
\usetikzlibrary{decorations.pathreplacing}
\usetikzlibrary{fit}
\usepackage{xspace}
\usepackage[table]{xcolor}

\crefname{section}{Sec.}{Sec.}
\crefname{figure}{Fig.}{Fig.}

\newcommand{\privkey}[1]{$k^{-}_{#1}$}
\newcommand{\pubkey}[1]{$k^{+}_{#1}$}

\newcommand{\code}[1]{\texttt{#1}}

\newcommand{\eg}{e.g.,\xspace}
\newcommand{\ie}{i.e.,\xspace}
\newcommand{\contribution}[1]{\textbf{#1}}
\newcommand{\goal}[1]{\textbf{\emph{#1}}}

\newcommand{\btcclient}{\code{basic-btc-client}\xspace}
\newcommand{\btcfirstversion}{Bitcoin Core~\code{26.0}\xspace}
\newcommand{\btcdefaultversion}{Bitcoin Core~\code{27.0}\xspace}
\newcommand{\btcversion}{Bitcoin Core~\code{28.0}\xspace}
\newcommand{\defaultport}{$8333$\xspace}
\newcommand{\relevantbip}{\gls{BIP}~324\xspace}
\newcommand{\measurementsstart}{24 May 2025\xspace}
\newcommand{\measurementsend}{7 July 2025\xspace}
\newcommand{\numin}{99513459}
\newcommand{\numout}{98971613}
\newcommand{\measurementsnuminmsgs}{\num{\numin}~\si{\million}\xspace}
\newcommand{\measurementsnumoutmsgs}{\num{\numout}~\si{\million}\xspace}
\newcommand{\measurementpps}{$\the\numexpr (\numin + \numout)/ (14*86400)pps$}
\newcommand{\micros}[1]{#1\si{\micro\second}}
\newcommand{\netfilter}{\code{netfilter}\xspace}
\newcommand{\testbed}{\code{btc-play-ground}\xspace}
\newcommand{\testbedurl}{\url{https://github.com/tud-dud/btc-play-ground}}
\newcommand{\pocurl}{\url{https://github.com/tud-dud/btc-v2-attack-tools}}
\newcommand{\simurl}{\url{https://github.com/tud-dud/btc-addr-sim}}
\newcommand{\syn}{\code{SYN}}
\newcommand{\tcpip}{TCP/IP\xspace}

\newcounter{measuresCtr}
\newcounter{issuesCtr}
\newcounter{subMeasureCtr}[measuresCtr]

\newcommand{\newissue}[3][1]{
\begin{enumerate}[\bfseries P1]
    \setcounter{enumi}{\value{issuesCtr}}
    \ifstrempty{#1}%
    {
        \item \textbf{#2:}#3
    }{
        \item \textbf{#2:}#3 \label{issue:#1}
    }
    \setcounter{issuesCtr}{\value{enumi}}
\end{enumerate}
}

\newcommand{\newmeasure}[2][1]{
\begin{enumerate}[\bfseries C1]
    \setcounter{enumi}{\value{measuresCtr}}
    \ifstrempty{#1}%
    {
        \item #2
    }{
        \item #2 \label{ctr:#1}
    }
    \setcounter{measuresCtr}{\value{enumi}}
\end{enumerate}
}

\newcommand{\newsubmeasure}[3][1]{
\edef\savedmeasure{\number\numexpr\value{measuresCtr}+1\relax}
\begin{enumerate}
    \renewcommand{\labelenumi}{\bfseries C\savedmeasure\Alph{enumi}}
    \setcounter{enumi}{0}
    \ifstrempty{#1}%
    {
        \item #2
        \item #3
    }{
        \item #1
        \item #2
        \item #3
    }
\end{enumerate}
\stepcounter{measuresCtr}%
}

\makeglossaries

\newabbreviation{AS}{AS}{autonomous system}
\newabbreviation{BGP}{BGP}{Border Gateway Protocol}
\newabbreviation{BIP}{BIP}{Bitcoin Improvement Proposal}
\newabbreviation{ECDH}{ECDH}{elliptic-curve-Diffie-Hellman}
\newabbreviation{HKDF-SHA256}{HKDF-SHA256}{HKDF-SHA256}
\newacronym{IPFS}{IPFS}{InterPlanetary File System}
\newabbreviation{LN}{LN}{Lightning network}
\newabbreviation{MITM}{MITM}{man-in-the-middle}
\newabbreviation{P2P}{P2P}{peer-to-peer}
\newacronym{TCP}{TCP}{Transmission Control Protocol}
\newacronym{V1}{V1 P2P transport}{Version 1 P2P transport}
\newacronym{V2}{V2 P2P transport}{Version 2 P2P transport}

\begin{document}

\maketitle
\nolinenumbers

\begin{abstract}
    Bitcoin recently introduced a new protocol for the encryption of
    \gls{P2P} communication.
    The protocol, known as \gls{V2}, represents a big step towards securing the
    overlay network against various previously-known attack vectors.
    Based on an analysis of \gls{V2}, this work examines the current viability
    of said attacks and concludes that while they are now remediated,
    alternative attacks and paths to similar objectives exist.
    The identified shortcomings are conceptual (and not implementation bugs) and
    even applicable to other \gls{P2P} networks.
    We show how a network-level attacker
    can identify application messages using the length of
    \glsentryshort{TCP} payloads, can eclipse a target node by taking advantage
    of how encrypted communication channels work and can downgrade all of a
    node's connections to the unencrypted protocol by using the mechanisms
    designed for compatibility.
    We validate our contributions using a combination of network measurements,
    emulations and simulations.
    Finally, we propose a series of short-term and long-term countermeasures
    towards securing Bitcoin's \gls{P2P} network.
    To the best of our knowledge, we are the first to study Bitcoin's security
    under \gls{V2}.
\end{abstract}

\glsresetall

\glsdisablehyper

\section{Introduction}
\label{sec:intro}

Despite its emphasis on decentralisation and security,
Bitcoin~\cite{nakamoto2008bitcoin} has long relied on
unencrypted \gls{P2P} communication.
Until recently, all network traffic between nodes
was exchanged over regular \gls{TCP} connections.
This proved to be a significant shortcoming of the Bitcoin protocol as multiple
attacks that were built on inspecting the plaintext communication were
uncovered.
Among others, there has been research on attacks on the
anonymity of users~\cite{biryukov2014deanonymisation, koshy2014analysis}, the
privacy of nodes in the overlay network~\cite{neudecker2016timing} and on the
security of \gls{P2P} communication in the presence of a malicious
\gls{AS}~\cite{apostolaki2021perimeter, apostolaki2017hijacking}.

As a result,
\relevantbip
defines a new \gls{P2P} transport
protocol, commonly referred to as \gls{V2}, which allows nodes to communicate
over encrypted but unauthenticated channels.
As of the release of \btcdefaultversion in April $2024$, \gls{V2} is the default
mode of communication.
Although it is reasonable to expect that \gls{V2} advances the security of the \gls{P2P}
network, the degree and consequences of these improvements are less clear.
To the best of our knowledge, neither the \gls{V2} protocol itself nor its
impact on the overlay's security have been studied.

In this paper, we present an analysis of Bitcoin's \gls{V2} protocol, with a
focus on known attacks by an \gls{AS}-level adversary.
We find that while encryption renders these attacks impractical in their
original form, an alternative path to them exists.
We first present an eclipse attack that exploits how \gls{V2} handles decryption
errors to close legitimate connections.
This replaces the component of prior attacks that has been disrupted by
\gls{V2}.
The issues are conceptual in nature and, to some extent, can be applied to other
major \gls{P2P} networks.
We also show that design choices made in the interest of backwards
compatibility, \ie silent retry and fallback on failure, give an adversary a
reliable way of downgrading connections to the unencrypted protocol.
Based on the insights from our analysis, we propose countermeasures to address
the identified weaknesses.
The main contributions of our work are summarised in the following.

\contribution{Analysis and Understanding of the V2 Protocol:}
We contribute to the understanding and further documentation
of the \gls{V2} protocol.
The latter is currently limited to the specification document, the reference
client's code base and a handful of informal resources.
Our work consolidates this information
with the intention of providing an additional source of reference.
For instance, we dissect the structure of a protocol message and show how message
types can be classified at the network layer.
We derive a threat model from \relevantbip, which we use for the rest of this
work, and discuss the current state of relevant, past attacks.
To the best of our knowledge, this is the first work to study the security of
Bitcoin's \gls{P2P} overlay network under \gls{V2}.

\contribution{Sustained Eclipse Attack by Exploiting Encryption:}
We present an effective network-level eclipse attack that does not require any
protocol state and replaces the components of previous
attacks~\cite{apostolaki2017hijacking, fan2021conman, tran2020stealthier} that
were broken by \gls{V2}.
The attack exploits how the protocol handles duplicate ciphertexts,
\ie a node closes the \gls{TCP} connection if it encounters
decryption errors.
We performed a review of how other major \gls{P2P} networks
handle decryption errors and found
that they all close the \gls{TCP} connections immediately.
This suggests that, although immediate disconnection upon decryption failures is
common practice, its use in \gls{P2P} environments
(as opposed to client–server architectures)
introduces a conceptual vulnerability that may be exploitable in other networks.

As a result of this design choice, an active network-level adversary can cause
all connections to a victim node to be closed by replaying payloads.
They then gradually occupy all connection slots until the victim node is eclipsed.
Furthermore, the attack exploits the fact that network-level classification of
message types in \gls{TCP} payloads is possible despite encryption.
Consequently, the adversary can replay payloads transporting the same message
type in order to maintain some discretion.
We implemented
the attack in our testbed and successfully eclipsed a victim node in less than a
day.
Our approach yields a more effective attack than the (now impractical) EREBUS
attack with a significantly shorter execution period as the adversary actively
provokes the closing of connections.
Although eclipse attacks in Bitcoin have been studied extensively~%
\cite{apostolaki2017hijacking, fan2021conman, heilman2015eclipse,
tran2020stealthier}, prior work neglects how to keep up the attack.
We therefore place emphasis on sustaining an eclipse attack and describe the
associated challenges and how they can each be overcome.

\contribution{Downgrade Attack by Exploiting Protocol Compatibility:}
We question the means with which \emph{compatibility} with the original
transport protocol, which we refer to as \gls{V1}, is achieved.
\relevantbip specifies that clients using \gls{V2} should accept unencrypted
incoming connections in order to minimise the risk of network partitions.
Compatibility is achieved by reconnecting using the \gls{V1} protocol if a
connection is terminated by the remote side immediately after the \gls{TCP}
handshake.
The issue therein is that \relevantbip relies on the transport layer for
protocol negotiation with older clients instead of making the necessary
provisions at the application layer.

We show how a network-level adversary can downgrade all of a node's connections
even when both nodes support \gls{V2} as a result of this design choice.
While the possibility of downgrade attacks is mentioned briefly in \relevantbip,
we are, to the best of our knowledge, the first to present such an attack and
demonstrate its feasibility.
Our analysis of \gls{V2} reveals a generous window of opportunity such that a
network attacker has sufficient time to trigger connection termination.
We implemented and confirmed that the attack is viable and succeeds without
fail.
This reopens the door to attacks that should no longer be possible due to
encryption such as delay~\cite{apostolaki2017hijacking} and
spoofing~\cite{fan2021conman, zou2024unveiling} attacks.

\contribution{Addressing Causes and Countermeasures:}
As our final contribution, we narrow down the causes of the various issues
identified in this work and suggest countermeasures Bitcoin can implement to
address them.
For example, certain features of the Bitcoin protocol, regardless of
\relevantbip, are conducive for eclipse attacks.
Some of the proposed countermeasures are short-term and not ample on their own,
while others address fundamental challenges in securing the \gls{P2P} network
and require further research.

\section{Bitcoin's \gls{V2} protocol}
\label{sec:p2p_layer}

Prior to December~$2023$, \gls{P2P} communication in the Bitcoin network was
not encrypted, which proved to have a significant impact on the
privacy~\cite{biryukov2014deanonymisation, koshy2014analysis} and
security~\cite{apostolaki2017hijacking, tran2020stealthier} of the \gls{P2P}
network.
The first notable attempts to address the lack of encryption were made in
\gls{BIP}~$151$~\cite{bip0151} in $2016$ which was later withdrawn in $2021$.
The specification of a new \gls{P2P} transport protocol in
\relevantbip~\cite{bip0324} and subsequent implementation in \btcfirstversion
introduced encrypted \gls{P2P} communication to Bitcoin.
This protocol, referred to as \gls{V2}, specifies a scheme for the encryption of
communication between peers and is enabled by default as of \btcdefaultversion.
Therefore, all \gls{P2P} communication subsequent to connection establishment
and key exchange will be encrypted if both peers support \gls{V2}.
In the following, we provide some background on peer and address management
followed by a description of the \gls{V2} protocol based on \relevantbip and the
\btcversion code base.
We then define a threat model based on our review of \gls{V2} and discuss the
current practicality of past attacks.

\subsection{Background: peer and address management}
\label{subsec:p2p}

The Bitcoin network is a permissionless \gls{P2P} network.
It primarily serves the purpose of propagating new transactions, blocks and node
addresses to the network via a gossip protocol.
Each node maintains up to $125$ connections by default.
Ten of these connection slots are reserved for outgoing connections and are
selected meticulously in order to ensure that a node is connected to at least
one honest node.
All but one of the remaining slots are reserved for incoming connections.
If a node already has the maximum number of open incoming connections when it
receives a further connection request, an eviction policy will determine whether
an existing connection should be closed in favour of the new one~\cite{ha2023on,
yang2023eviction}.

Addresses discovered during bootstrapping and thereafter are managed in a local
database that is is divided into two tables.
The \emph{new} table stores
addresses that are relayed via
\code{ADDR} messages, while
the \emph{tried} table stores
addresses of nodes that have been successfully connected to.
Addresses are organised into \emph{buckets} primarily based on the $/16$ or
$/32$ prefix of an IPv4 or IPv6 address respectively and the relaying node's
prefix group.
Each address in the new table can occur in up to 8 different buckets.

The final connection slot is reserved for a short-lived outgoing connection, a
\emph{feeler} connection, which is initiated every two minutes to a random
address from the new table.
The point of the feeler connection is to establish if a given address in the new
table is reachable.
If that is the case, the address is moved to the tried table.
When initiating a new outbound connection, a node first chooses either the new
or tried table with equal probability.
It then selects a random address from the chosen table and attempts to connect
to it.
If the connection is established, the address is moved to the tried table in
case it was selected from the new table.
Any existing copies of the address in the new table are also deleted.

\subsection{\gls{V2} protocol}
\label{subsec:v2}

\relevantbip defines three components of the \gls{V2} protocol: the
\emph{transport layer}, the \emph{application layer} and the \emph{signalling}
components.
These subprotocols collectively contribute to a total of eight objectives
defined in \relevantbip.
We provide an overview of the \gls{V2} protocol and highlight its objectives in
what follows.
We concentrate on the establishment of encrypted communication channels.
All of the objectives are named and highlighted in bold italics below, and a
summary is provided in \cref{tab:v2_goals} in the appendix.

\subparagraph{V2 Transport layer}

\begin{figure}
    \begin{minipage}[b]{.45\textwidth}
    \scalebox{0.6}{
    \begin{tikzpicture}
        \tikzstyle{label}=[above,midway,sloped,font=\footnotesize]

        \draw[stealth-,thick] (-0.9, -6.5) -- node[above,midway,sloped,]{\small
        Time}(-0.9, 0.25);
        \path
            (0,0) node[draw] (i) {Initiator}
            (6,0) node[draw] (r) {Responder};
            \draw[thick] (i) -- ++(0,-6.5) (r) -- ++(0,-6.5);

        \node[] at (-0.25,-0.7) {\privkey{I}};
        \node[] at (6.25,-0.7) {\privkey{R}};

        \draw[->,thick] (0, -1) -- node[label]{\pubkey{I}}(6,-1);
        \draw[->,thick] (0, -1.5) -- node[label]{\emph{garbage}}(6,-1.5);
        \draw[<-,thick] (0, -2.0) -- node[label]{\pubkey{R}}(6,-2.0);
        \draw[<-,thick] (0, -2.5) -- node[label]{\emph{garbage}}(6,-2.5);
        \draw[->,thick] (0, -3.0) -- node[label]{\emph{garbage terminator}}(6,-3);
        \draw[<-,thick] (0, -3.5) -- node[label]{\emph{garbage terminator}}(6,-3.5);
        \draw
        [decorate,decoration={brace,amplitude=4pt},xshift=0.4cm,yshift=0pt]
        (6.25, -0.5) -- (6.25, -3.5) node [midway,right,xshift=.1cm,align=left]
        {Key\\exchange};

        \draw[->,thick] (0, -4.0) -- node[label]{\code{VERSION}}(6,-4);
        \draw[<-,thick] (0, -4.5) -- node[label]{\code{VERSION}}(6,-4.5);
        \draw
        [decorate,decoration={brace,amplitude=4pt},xshift=0.4cm,yshift=0pt]
        (6.25, -3.5) -- (6.25, -5.0) node [midway,right,xshift=.1cm,align=left]
        {Version\\negotiation};

        \draw[->,thick] (0, -5.0) -- node[label]{\code{VERSION}}(6,-5);
        \draw[<-,thick] (0, -5.5) -- node[label]{\code{VERACK}}(6,-5.5);
        \draw[<-,thick] (0, -6) -- node[label]{\code{VERSION}}(6,-6);
        \draw[->,thick] (0, -6.5) -- node[label]{\code{VERACK}}(6,-6.5);
        \node[align=center] at (3, -6.75) {$\vdots$};
        \draw
        [decorate,decoration={brace,amplitude=4pt},xshift=0.4cm,yshift=0pt]
        (6.25, -5.0) -- (6.25, -7) node [midway,right,xshift=.1cm,align=left]
        {Application~(same in\\\gls{V1}\\but w/o encryption)};

    \end{tikzpicture}
    }
    \caption{\gls{V2} connection establishment after the \gls{TCP}
        handshake.
    }
    \label{fig:v2_handshake}
    \end{minipage}
    \hfill
    \begin{minipage}[b]{.45\textwidth}
    \scalebox{0.7}{
    \begin{tikzpicture}
        \tikzstyle{label}=[above,midway,sloped,font=\footnotesize]

        \node[draw, rectangle,align=center] (ld) {Length descriptor\\$3$~bytes};;
        \node[draw, rectangle,align=center,right=of ld,xshift=-2.75em] (ct)
        {Ciphertext\\$n$~bytes};
        \node[draw, rectangle,align=center,below=of ct,xshift=-2em,yshift=1.5em] (h) {Flags\\$1$~byte};
        \node[draw, rectangle,align=center,right=of h,xshift=-2.75em] (content)
        {Contents\\$n-17$~bytes};
        \node[draw, rectangle,align=center,right=of content,xshift=-2.75em]
        (tag) {Auth. tag\\$16$~bytes};
        \node[draw, rectangle,align=center,below=of
        content,xshift=-2em,yshift=1.5em] (t)
        {Type\\$1$ or $3$~bytes};
        \node[draw, rectangle,align=center,right=of t,xshift=-2.75em]
        (msg) {Message\\$n-18$ or $n-20$~bytes};

        \draw[-] (ct) -- (h);
        \draw[-] (ct) -- (content);
        \draw[-] (ct) -- (tag.north);

        \draw[-] (content) -- (t);
        \draw[-] (content) -- (msg);

    \end{tikzpicture}
    }
    \caption{Structure of a \gls{V2} message during the application phase.
    }
    \label{fig:v2_packet}
    \end{minipage}
\end{figure}

The transport layer defines three phases~(cf. \cref{fig:v2_handshake}) and is
responsible for setting up an encrypted channel over a \tcpip connection between
two nodes and encrypting protocol messages.
The channel is then used to transport application messages.

In the \emph{key exchange phase}, the initiator and responder
each generate an ephemeral secp256k1
private key, \privkey{I} and \privkey{R} respectively, and exchange the
corresponding 64-byte public keys (\pubkey{I} and \pubkey{R}).
Both parties use X-only \glsentryshort{ECDH}~\cite{bernstein2006curve} to
compute a shared secret from the various keys and derive four encryption keys.
They each derive one key for the encryption of packet lengths and
one each for content encryption.
They also derive a common session ID and a 16-byte garbage terminator each using
\glsentryshort{HKDF-SHA256}~\cite{krawczyk2010cryptographic}.
The session ID provides \goal{observability of active attacks} as peers can
detect an active \gls{MITM} attacker by comparing them manually.
The garbage terminators are then exchanged and all further communication
takes place in the form of encrypted packets leading
to \goal{confidentiality against passive attacks}.
The protocol's main \goal{shapability} mechanism is provided by decoy packets
that may be sent at any point henceforth.
However, \relevantbip does not specify how decoy packets are to be used.

The \emph{version negotiation phase} serves the purpose of peers agreeing on a
transport protocol version to use.
Support for the current \gls{V2} protocol is signalled by each node sending an
empty \code{VERSION} packet.
This is distinguishable from the original handshake because empty \code{VERSION}
messages are never sent.
Protocol \goal{upgradability} is achieved in this way as future versions will be
advertised via the content of this \code{VERSION} message.

During the \emph{application phase}, data transmitted between nodes in network
packets is interpreted as application data.
These packets contain application messages (commonly referred to as
\emph{contents}).
The application phase is the same in the case of either transport protocols and
always starts with a mandatory exchange of the \code{VERSION} and \code{VERACK}
messages.
These messages are used to communicate the features a node supports.
\gls{V2} maintains \goal{low overhead} compared to the \gls{V1} protocol
and even a slight bandwidth reduction due to, \eg smaller application messages.

\subparagraph{V2 Application layer}

The application layer is responsible for encoding application messages for
transport by the transport layer.
Each transmitted packet is authenticated and encrypted.
Authentication in the scope of \relevantbip refers to the guarantee that a
message obtained via successful decryption was encrypted using the legitimate
encryption key.
As depicted in
\cref{fig:v2_packet},
every \gls{V2} packet contains an encrypted $3$-byte length descriptor and a
variable-length ciphertext consisting of the following fields:
\begin{itemize}
    \item a $1$-byte header for protocol flags.
        Currently, only the highest bit is defined as the ignore
        bit for decoy packets;
    \item a $16$-byte Poly1305 authentication tag of the encrypted plaintext;
        and
    \item a variable-length array of the application message which is commonly
        referred to as the \emph{contents}.
        The length is defined to be in the interval $[0, 2^{24}-1]$ and is
        encoded in the aforementioned $3$-byte length descriptor.
\end{itemize}
The \emph{contents} field consists of:
\begin{itemize}
    \item a $1$-byte field encoding the message type; and
    \item a variable-length byte array of the actual message.
\end{itemize}
The plaintext is encrypted using ChaCha20-Poly1305~\cite{rfc8439}, while the
length descriptor is encrypted using the ChaCha20 block function~\cite{rfc8439} with
a different key.
Consequently, each \gls{V2} \gls{TCP} payload will have a minimum size of $20$
bytes and a maximum size of $2^{24}+19$ bytes.
Encrypting the fixed-size length field guarantees a stream that is
\goal{pseudorandom} to a passive attacker, \ie they should not be able to
distinguish a \gls{V2} bytestream from a uniformly random bytestream.
Both encryption keys are rotated every $224$ packets in order to provide
\goal{forward secrecy} in case either is compromised.

\subparagraph{V2 Signalling}

The signalling component defines how nodes announce support for the \gls{V2}
protocol to other nodes in the network.
They do so by advertising the \code{NODE\_P2P\_V2} service flag when announcing
their address to the network.
For the sake of \goal{compatibility}, client implementations are encouraged to
attempt connecting with the V1 protocol if they are met with immediate
disconnection during the handshake phase.

\subsection{Threat model}
\label{subsec:threat}

\relevantbip does not explicitly define a threat model.
Yet, it is evident that \gls{V2} is designed to protect nodes from network-level
attackers.
Such attackers can vary both in terms of their capabilities, \eg an
eavesdropper within a network or an \gls{AS}, as well as in their methodology,
\ie active or passive.
We therefore describe our threat model in what follows based on related work and
conclusions we draw from \relevantbip.

This work assumes a network-level adversary, \eg an \gls{AS}, that has control
of key network infrastructure between a victim node and the rest of the
Bitcoin network.
This is a common threat model in the context of blockchain-based \gls{P2P}
networks.
Especially in the case of Bitcoin, multiple works have studied the network's
security in the presence of such an adversary~\cite{apostolaki2021perimeter,
apostolaki2017hijacking, fan2021conman, tran2020stealthier}.
This threat model has also been applied to the
\glsentrylong{LN}~\cite{ndolo2024payment, arx2023revelio} and related
networks~\cite{saad2023three}.
Depending on the specific attack, the adversary is either passive or active.
In the latter case, they are in-path~\cite{marczak2015analysis} and are thus
capable of tampering with \gls{TCP} streams arbitrarily, \eg by dropping packets
to/from the victim or by injecting forged packets into the stream.
Even if the adversary is not initially in-path, techniques such as BGP hijacking
and interception attacks provide a viable route to that
position~\cite{apostolaki2017hijacking, birge-lee2019sico, birge-lee2025global}.
Furthermore, our eclipse attack
requires that the adversary is able to identify the type of application message
in an encrypted \gls{TCP} payload.
While the message format did not change in \gls{V2}, message types can
no longer be identified by reading the payload.
We assume that message types can still be inferred
from side-channel information, in particular message length,
which we explain in \cref{subsec:classification}.

We require that potential victim nodes are running \btcdefaultversion or higher
and that all of their connections are using the \gls{V2} protocol.
As in previous work, the adversary is only interested in nodes connecting to the
network directly via \tcpip, and not via Tor.
Moreover, as \gls{V2} is not self-revealing, unlike V1, we assume that potential
victim nodes are using the default port~(\gls{TCP}/\defaultport) for \gls{P2P}
communication.

\subsection{Network attacks under \gls{V1}}
\label{subsec:historical_attacks}

The first class of attacks of interest is eclipse and partitioning attacks as
they have been of continued interest in Bitcoin.
The first eclipse attack in Bitcoin was presented in
2015~\cite{heilman2015eclipse} at a time when the Bitcoin protocol had several
shortcomings in its address management.
The adversary model later shifted to consider powerful network attackers who
have access to and are capable of manipulating routing
infrastructure~\cite{apostolaki2017hijacking, fan2021conman, ha2023on,
tran2020stealthier}.
These attacks exploit the properties of the communication protocols used in the
Internet as well as Bitcoin's pre-\gls{V2} lack of encryption and/or
authentication to eclipse a victim node.
\relevantbip successfully renders these attacks infeasible: encryption prevents
an adversary from reading or modifying \gls{TCP} payloads as well as from
spoofing.
\relevantbip hinders any interference with existing connections but it does not
change address management in Bitcoin.

The second class of attacks comprises the remaining network-level attacks
identified in the literature.
Their goals vary considerably and range from delaying block
propagation~\cite{apostolaki2017hijacking} to
off-path spoofing~\cite{li2023bijack}.
The attacks all exploit the absence of encryption and are therefore no longer
practical threats as an adversary cannot read or modify the content of the
payloads.
While they could attempt traffic fingerprinting as described in
\cref{subsec:classification},
it is potentially less reliable and requires more resources.

\subsection{Network-level message classification}
\label{subsec:classification}

Our threat model assumes that it is possible to determine the type of Bitcoin
messages in a \gls{TCP} payload using the payload's length.
\gls{V2} does not specify any shapability mechanism despite it being one of the
protocol's objectives.
The ability to identify message types despite encryption
introduces a side channel for other attacks.

Given a \gls{TCP} payload with $x$ bytes, we can infer that it contains an
application message of type $t$ if subtracting $20$~(mandatory fields) and
either $1$ or $3$~(type field) from $x$ returns a multiple of $t$'s minimum
length, \ie if $(x-21-\{1,3\}) \bmod min(\left| t \right|) = 0$.
We validated this methodology using a $6$-week measurement study of the public
network and evaluated its merit.
While certain messages types, \eg \code{VERSION}, can be identified reliably,
other message types, \eg \code{BLOCK}, cannot be identified using an equally
straightforward approach due to several variable-length fields.
Furthermore, as application messages are transmitted as continuous streams of data
over \gls{TCP}, a payload may carry more than one application message.
We expound on this approach in \cref{app:classification} and discuss the study
in \cref{sec:measurements}.

\section{Eclipse attack}
\label{sec:partitioning}

Partitioning attacks isolate a set of nodes from the rest of a network by
splitting the network.
Eclipse attacks isolate a single node from the rest of the network, and are
considered to be a subclass of partitioning attacks~\cite{franzoni2022sok}.
The main motivation for such attacks is to control the flow of legitimate
information to victim nodes.
In doing so, they become vulnerable to further attacks that are usually
mitigated by the consensus rules in the network, \eg
double-spending~\cite{karame2012double} or $51\%$
attacks~\cite{nakamoto2008bitcoin}.

Eclipse attacks under \gls{V2} require a new mechanism for connection
manipulation that does not depend on payload inspection or spoofing.
A naive attack by means of dropping specific timeout-relevant messages results
in immediate disconnection as decryption of any subsequent payload fails.
Based on this observation, we present a new approach to closing the victim's
existing connections in order to free up connection slots.
As the general approach to occupying connection slots described
in~\cite{fan2021conman, tran2020stealthier} still applies, we place greater
emphasis on how to sustain an eclipse attack.
We provide new insights into challenges an adversary will face and how each can
be overcome.
To the best of our knowledge, we are the first to provide an elaborate
description of this component of eclipse attacks.

\subsection{Closing existing connections}

Our attack leverages that \gls{V2} connections are terminated as a consequence
of malleability.
Specifically, the \emph{in-path} adversary triggers
disconnection by duplicating specific \gls{TCP} payloads from the victim to its
peers, \ie a replay attack.
As \gls{V2} ciphertexts can only be decrypted exactly once, recipients of
such ciphertexts will not be able to decrypt them and assume a
transport layer error.
As a result, they close the \gls{TCP} connection using a regular \code{FIN}
packet, which opens up a connection slot for the adversary.
The adversary performs the following steps in order to eclipse a victim node
successfully:
\begin{enumerate}
    \item The adversary monitors all traffic to/from the victim on \gls{TCP}
        port \defaultport and makes decisions on how packets should traverse the
        network stack.
        All packets are sent to their destination without any form
        of modification by default.
    \item The adversary selects an application message type $m$ that they will
        use for the attack.
        As the \gls{P2P} traffic is encrypted, the adversary does not know what
        message types are being transmitted.
        They use the size of \gls{TCP} payloads to determine the type
        (cf.~\cref{subsec:classification}).
    \item The adversary waits for the first packet carrying a message of type
        $m$ from the victim node to a given peer.
        They store a copy of the payload and dispatch the packet as it is.
    \item When the victim node sends a second packet with a message of type $m$,
        the adversary replaces the packet's payload with the previously stored
        copy of $m$.
        They recompute the necessary fields of the \gls{TCP} header before
        letting it traverse the rest of the network stack regularly.
        In order to reduce the risk of detection and minimise their workload,
        they only modify packets with messages of the same size and type.
    \item When the altered packet arrives at its destination, the application
        will decrypt the $3$-byte length descriptor which will return a wrong
        length.
        This is because the ChaCha20 block counter will have advanced and the
        keystream's bytes used for decryption cannot produce the
        correct length.
        A \enquote{packet too large} error message will be written to the
        recipients logs before closing the connection to the victim
        using a regular four-way handshake of \gls{TCP} \code{FIN}s.
        This leaves no indications of an attack.
        Unlike a classic transport layer attack, \eg \code{RST} injection,
        failed decryption is far less indicative of in-path manipulation.
    \item Incoming peers will try to reconnect to the victim node which the
        attacker must prevent by dropping the \gls{TCP} \syn\xspace packets.
        The attacker occupies every newly-available incoming connection slot
        with a malicious connection.
        In the case of outbound connections from the victim node, the adversary
        must block any subsequent reconnection attempts.
        They, however, have no other choice but to wait for the victim to open
        connections to them.
\end{enumerate}

The adversary follows these steps for every honest peer that the victim has
until no more persist.
The strategy does not entail preventing connections from new peers as they will
simply be compromised later.
We remark that the applicability of the attack is not exclusive to Bitcoin.
A review of related \gls{P2P} networks such as \gls{IPFS}, the \glsentrylong{LN}
and Ethereum showed that it is common practice to close \gls{TCP} connections
when decryption fails.
This creates a broader conceptual vulnerability that could be exploited in other
networks.

\subsection{Choice of message to duplicate}

Technically, any application message type that is sent more than once to each
peer can be used for the attack.
However, certain properties reinforce the attack from an adversarial point of
view.
The more often the victim sends a certain message type of the same size, the
more suitable it is for the attack for two main reasons.
On the one hand, connections can be closed faster and on the other, the attacker
only needs to store payloads for a short time.

We therefore recommend using either the \code{PING} or \code{PONG} message
types.
\code{PING}s are sent approximately every two
minutes~\cite{bitcoincorepings} and are the smallest non-empty messages.
However, both messages are transported in $29$-byte payloads which makes them
indistinguishable based on just the payload size.
As the adversary does not maintain any protocol state, they will inevitably be
wrong sometimes, \ie \code{PING}s may be replaced with \code{PONG}s or vice
versa.

\subsection{Occupying the target's connection slots}
\label{subsec:outgoing}

A single honest connection to the network is sufficient for a node to
remain in consensus with the rest of the network.
A successful eclipse attack therefore requires that each of the victim's peers
is controlled by the adversary.
Inbound connection slots can be filled easily by the adversary as there are no
requirements for them.
Hence, it suffices to connect to the target from one host with \emph{one} IP
address on different ports until all slots are occupied.

Filling the outgoing connection slots is significantly more challenging as there
is not much the adversary can actively do besides making sure that their addresses are
inserted into the new table. %
The approach described in~\cite{tran2020stealthier} still works even though a
series of measures was deployed in response to previous eclipse
attacks~\cite{apostolaki2017hijacking, heilman2015eclipse, tran2020stealthier}.
We therefore provide a pertinent overview of the steps an adversary must take in
order to receive connections from the victim; a detailed description is
available in
\cref{subsec:outgoing_connections}.
\begin{enumerate}
    \item The adversarial addresses must be inserted into the victim node's
        address database.
        The adversary inserts its addresses by participating in the \gls{P2P}
        network and relaying its addresses via regular \code{ADDR} to the victim
        node;
    \item the adversary must have at least \emph{ten} IP addresses from distinct
        $/16$ IPv4 or $/32$ IPv6 subnets.
        As our threat model assumes a network-level adversary such as an
        \gls{AS}, we assume that they are able to meet this requirement;
    \item the size of the address table means that the probability of selection
        is low.
        The adversary improves their chances by ensuring that each address occurs
        multiple times in the new table and that they are all from different prefix
        groups; and
    \item after an address has been selected, Bitcoin Core performs various
        checks to verify that the address is suitable as an outgoing
        connection.
        Among others, they check that the selected node supports a series of
        application-level features.
        Adversarial nodes must announce each of these features, however, they do
        not need to actually support any of them.
\end{enumerate}

\subsection{Maintaining connections}
\label{sub:sustaining_attack}

The eclipse attack is successful once the adversary has managed to occupy all of
the victim node's connection slots.
Depending on the adversary's ultimate objective, the adversary needs to sustain
the connections for follow-up attacks.
Bitcoin core implements various mechanisms to make sure that connections are
active and useful to a node.
Connections that do not fulfil these requirements are replaced.
We identified the minimum functionality a client must implement in order
to maintain the \gls{P2P} connections.
In the following, we provide an overview of the relevant mechanisms and how each
can be overcome.
\btcclient is our implementation of such a client and is described in
\cref{subsec:impl_eclipse}.
It implements the necessary functionality to bypass the challenges while
maintaining no protocol state.

\subparagraph{Application layer handshake}

Subsequent to the \gls{TCP} handshake, a node will wait for up
to $60$ seconds for the initiator to start the \gls{P2P} handshake
(cf.~\cref{fig:v2_handshake}).
If that does not happen, the newly-established connection will be closed.
Hence, the bare minimum an adversarial node must do to avoid immediate
disconnection is to perform the handshake with the victim.

\subparagraph{Timeout}

Several timeouts lead to disconnection.
However, with respect to a minimal \gls{P2P} peer that does not relay data, the
most relevant is a $20$-minute inactivity
timeout~\cite{bitcoincoretimeout}.
It expires if either a \code{PONG} in response to a \code{PING} does not arrive
in time or, more generally, if a node has not received any communication from a
peer within the timeout.
Adversarial nodes can avoid disconnection simply by responding to \code{PING}s
from the victim node.

\subparagraph{Synchronisation of outbound peers}

A node may disconnect an outbound peer if they are determined to be out-of-sync
with the node's blockchain, \ie if they have never sent a block header or have
only sent one with old information~\cite{bitcoincoresync}.
A node verifies that an outbound peer is synchronised by sending them a
\code{GETHEADERS} message requesting block headers from a particular point
in the blockchain.
The response, a \code{HEADERS} message, must be received within two minutes.
Failure to respond within the timeout or responding with stale data leads to
immediate disconnection.

As the adversarial nodes do not propagate chain data, they will be inevitably
asked to verify their chain state after $20$ minutes.
In order to pass the verification and keep the connection open, adversarial
nodes can keep track of the chain state by issuing \code{GETHEADERS} messages
periodically.
An alternative is to send the exact same \code{GETHEADERS} message back to the
victim, wait for their \code{HEADERS} reply, and then simply send the received
data back to the victim node.
The latter approach is preferable as it does not require protocol state but
relies on the victim's \code{HEADERS} message arriving before the two-minute
timeout elapses.

\subparagraph{Eviction of inbound peers}

The victim may still receive incoming connection requests from other nodes in
the network.
This triggers the eviction mechanism that determines which peer to disconnect so
as to free up a slot for the new connection.
While the eviction policy is complex and out-of-scope of this work, the
adversary is at a clear disadvantage for two reasons: they do not provide any
useful data to the victim and each of the incoming connections were initiated
from the same IP address.
We refer to~\cite{ha2023on, yang2023eviction} for
descriptions of the eviction mechanism in Bitcoin.

The adversary has two straightforward options of dealing with potential
eviction.
They can either block all incoming \gls{TCP} \syn{}s after eclipsing the victim
or allow such connections and risk temporary eviction before closing them as
part of the attack.

\section{Downgrade attack}
\label{sec:downgrade}

In the following, we present a downgrade attack that reopens the door to the
full range of legacy attacks.
The objective of the downgrade attack is to force the victim into using the
original unencrypted protocol for all \gls{P2P} communication.
The attack leverages the fact that \gls{V2} lacks a version negotiation
protocol and relies on the transport layer for signalling of compatibility.
The adversary can therefore first perform an additional step of downgrading the
victim's connections before proceeding to their primary attack.

\subsection{Downgrading \gls{V2} connections}

\begin{figure}
    \centering
    \scalebox{0.6}{
    \begin{tikzpicture}
        \tikzstyle{label}=[above,midway,sloped,font=\footnotesize]

        \draw[stealth-,thick] (-0.9, -5) -- node[above,midway,sloped,]{\small
        Time}(-0.9, 0.25);
        \path
            (0,0) node[draw] (i) {Initiator}
            (6,0) node[draw] (r) {Responder};
            \draw[thick] (i) -- ++(0,-5) (r) -- ++(0,-5);

        \node[] at (-0.25,-0.7) {\privkey{I}};
        \node[] at (6.25,-0.7) {\privkey{R}};

        \node[align=center] at (3, -0.3) {\emph{(\gls{TCP} handshake)}};
        \draw[->,thick] (0, -1) -- node[label]{\pubkey{I}}(6,-1);
        \draw[->,thick] (0, -1.5) -- node[label]{\emph{garbage}}(6,-1.5);
        \draw[<-,thick] (0, -2.0) -- node[label]{\pubkey{R}}(6,-2.0);
        \draw[<-,thick] (0, -2.5) -- node[label]{\emph{garbage}}(6,-2.5);
        \draw[->,thick] (0, -3.0) -- node[label]{\emph{garbage terminator}}(6,-3);
        \draw[<-,thick] (0, -3.5) -- node[label]{\emph{garbage terminator}}(6,-3.5);
        \draw
        [decorate,decoration={brace,amplitude=4pt},xshift=0.4cm,yshift=0pt]
        (6.25, -0.5) -- (6.25, -3.5) node [midway,right,xshift=.1cm,align=left]
        {Window of\\opportunity};

        \draw[->,thick] (0, -4.0) -- node[label]{\code{VERSION}}(6,-4);
        \draw[<-,thick] (0, -4.5) -- node[label]{\code{VERSION}}(6,-4.5);
        \node[align=center] at (3, -4.75) {$\vdots$};

    \end{tikzpicture}
    }
    \caption{
        Overview of the \gls{V2} protocol connection establishment scheme
        showing the window of opportunity for downgrade attacks.
        \code{RST} packets sent during this period will cause the initiator to
        retry using the V1 protocol.
}
    \label{fig:v1_downgrade}
\end{figure}
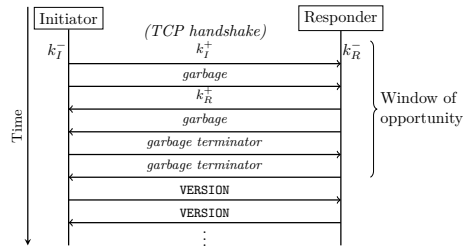

An active network-level adversary can take advantage of \relevantbip's
compatibility considerations to ensure that a victim node only uses the \gls{V1}
protocol.
In order to understand how they can do so, we revisit the different phases a new
\tcpip connection goes through.
As illustrated in
\cref{fig:v1_downgrade}, nodes exchange
keys and generate shared secrets after connection establishment.
\relevantbip encourages implementations to retry using the V1 protocol if they
are met with \emph{immediate disconnection} when establishing a \gls{V2}
connection.
While immediate is not specified, Bitcoin Core will retry if the responder
terminates the connection before completion of the key exchange phase.

Therefore, an adversary can downgrade the connection by injecting a \gls{TCP}
\code{RST} packet at any point during the key exchange phase.
They can do so by either changing a legitimate packet to a \code{RST} packet
(in-path~\cite{marczak2015analysis}) or injecting a forged \code{RST} packet
into the \gls{TCP} stream (on-path~\cite{weaver2009detecting}).
However, it has been shown that on-path attack systems can generally be detected
by observing anomalies in the traffic flows~\cite{marczak2015analysis,
weaver2009detecting}.
In order to comply with the protocol specification, \eg for the sake of
discretion, the \code{RST} packet should be sent from the responder to the
initiator.
However, a Bitcoin Core initiator will retry the connection regardless of where
the \code{RST} originated from as long as it is during the window of
opportunity shown in \cref{fig:v1_downgrade}.
The initiator will then open a new \gls{TCP} connection to the responder, skip
the key exchange and version negotiation phases, and initiate the regular
handshake of \code{VERSION}/\code{VERACK} messages.
Evidently, no communication over this connection will be encrypted.
The adversary ensures that all of a victim's connections use the V1
transport protocol by consistently interfering with all connection attempts
to/from the victim node.

\subsection{Identifying V2 connection attempts}

As described so far, the adversary interferes with all connection attempts
regardless of the transport protocol version the connection would use.
That would cause more damage than intended to the victim node because V1
transport connections will not be retried.
Our measurements in the mainnet showed that less than $50\%$ of the node's
outgoing connections are \gls{V2} connections.
Since the attack's objective is to ensure that all of the connections are
unencrypted, the adversary does not need to interfere with V1 connection
attempts.
The adversary, therefore, needs to be able to differentiate between connections
that will use \gls{V1} and \gls{V2}.

The protocol version a new connection will use can be determined based on
the payload of the first packet after the \gls{TCP} handshake.
The adversary can use either the content or the size of the first
payload to distinguish the protocol version that is being set up.
The initiator will use \gls{V2} if the \code{NODE\_P2P\_V2} service flag was
previously announced for the responder's address.
If a V1 connection is being established, the initiator's first packet after the
\gls{TCP} handshake will contain $126$ bytes of payload as it is a
\code{VERSION} message (cf.
\cref{fig:v2_handshake}).
In contrast, \gls{V2}'s first payload is larger and cannot be
serialised to \code{VERSION} message.
The adversary only needs to downgrade the connection in the latter
case.

\subsection{Window of opportunity}

\cref{fig:v1_downgrade} shows the window of opportunity for this attack.
It is the time period during which public keys, garbage and garbage terminators
are exchanged.
Bitcoin Core implements this phase in one and a half round trips as follows:
\begin{enumerate}
    \item initiator sends their \pubkey{I} and garbage;
    \item responder sends their \pubkey{R}, garbage, garbage terminator and
        \code{VERSION} message; and
    \item initiator sends their garbage terminator and \code{VERSION}.
\end{enumerate}
This makes the critical period shorter in practice and requires that the attack
is implemented efficiently.
Nonetheless, there remains ample time for a successful attack.

\subsection{Attack duration}
\label{subsec:downgrade_reboot}

The downgrade attack depends on the adversary's ability to interfere with
connections while they are being established.
If the victim node already has some peers, the attack will require a
significant amount of time as connections tend to be
kept alive for extended periods~\cite{apostolaki2017hijacking,
neudecker2019characterization}.
The attack can be sped up significantly if the adversary waits for or is able to
trigger a reboot of the victim node.
Besides deliberate efforts to restart a node such as DDoS, memory exhaustion or
power failures, there are more predictable reasons such as major software
upgrades~\cite{heilman2015eclipse}.
For instance, according to the Bitnodes crawler, $120$ nodes had upgraded to
\btcversion one day after its release on 4 October 2024.
The number was up to $447$ nodes a week later and $1562$ nodes after $30$ days.
In other words, there were at least $1562$ restarts in the first month after
\btcversion's release.
Furthermore, previous work~\cite{biryukov2014deanonymisation} found that a node
with a public IP address has a $10\%$ probability of going offline after two
hours. The probability increases to $25\%$ after ten hours.

\section{Evaluation: Eclipse Attack}
\label{sec:evaluation_eclipse}

In the following, we describe our evaluation setup before presenting results
concerning various aspects of the attack.
Multiple results are omitted due to space constraints but are provided in
\cref{fig:supp_eclipse} as part of the appendix.
The impact of eclipse attacks on Bitcoin has been evaluated extensively in
previous work (refer to~\cite{ha2023on} for an analysis of the impact of eclipse
attacks).
Our evaluation therefore places emphasis on the primary contributions of this
work.

\subsection{Methodology}
\label{subsec:methodology}

    \begin{figure}
    \centering
    \scalebox{0.45}{
        \begin{tikzpicture}[]
            \tikzstyle{label}=[midway,font=\huge,red]
            \tikzstyle{box}=[minimum]

            \node[draw,thick,rectangle,align=left,inner sep=10pt] (router) {Router};
            \node[draw,thick,rectangle,align=left, below left=of
            router, text width=10cm, minimum height=2.25cm] (regnet) {
              \btcdefaultversion or \btcversion
              \vspace{1ex}\hrule\vspace{1ex}
              $175.i.0.j/16,$
              $i \in [1, 200], j \in [1,3]$
            };

            \node[draw,thick,rectangle,align=left, below right=of router, text
            width=10cm, minimum height=2.25cm] (attnet) {
              \btcclient(listen mode)
              \vspace{1ex}\hrule\vspace{1ex}
              $11.i.0.1/16, i \in [1,20]$
            };

            \node[draw,thick,rectangle,align=left, above left=of router, text
            width=10cm, minimum height=2.25cm] (seednodes) {
              \btcversion
              \vspace{1ex}\hrule\vspace{1ex}
              $193.168.1.i/24, i \in [3, 5]$
            };

            \node[draw,thick,rectangle,align=left, above right=of
            router, text width=10cm, minimum height=2.25cm] (victim) {
              \btcversion
              \vspace{1ex}\hrule\vspace{1ex}
              \setlength{\fboxrule}{2pt}
              \fcolorbox{red}{white}{\parbox{0.95\textwidth}{
                \btcclient(connect mode)*
                \vspace{1ex}\hrule\vspace{1ex}
                \code{iptables} and
                \netfilter-based attack code
              }}
              \vspace{1ex}\hrule\vspace{1ex}
              $193.168.1.2/24$
            };

            \node[red,scale=2.5,anchor=east,yshift=15pt,xshift=3pt]
              at (victim.south west) {\faGhost};

            \draw[thick,-] (router) -- (regnet);
            \draw[thick,-] (router) -- (attnet);
            \draw[thick,-] (router) -- (seednodes);
            \draw[thick,-] (router) -- (victim);

            \node[above=1mm of regnet] (l0) {\textbf{Regular nodes}};
            \node[above=1mm of seednodes] (l1) {\textbf{Seed nodes}};
            \node[above=1mm of victim] (l2) {\textbf{Victim node}};
            \node[above=1mm of attnet,align=left] (l3) {\textbf{Attacker nodes}*};

            \node[anchor=north west,font=\LARGE] (note) at (attnet.south west)
              {* only for the eclipse attack};

            \node[fit={(router)(regnet)(seednodes)(victim)(l0)(l1)(l2)(l3)(note)}, draw,
            dashed, very thick, inner sep=10pt] (fit) {};
            \node[above right] at (fit.north west)
            {Virtual Mininet network};

            \node[red,anchor=north west] at (fit.south west) {\faGhost
              \hspace{1ex}\color{black} position simulates a network-level attacker};
        \end{tikzpicture}
        }
        \caption{Network topology used for the evaluation of the eclipse and
            downgrade attacks. %
        }
        \label{fig:topology}
    \end{figure}
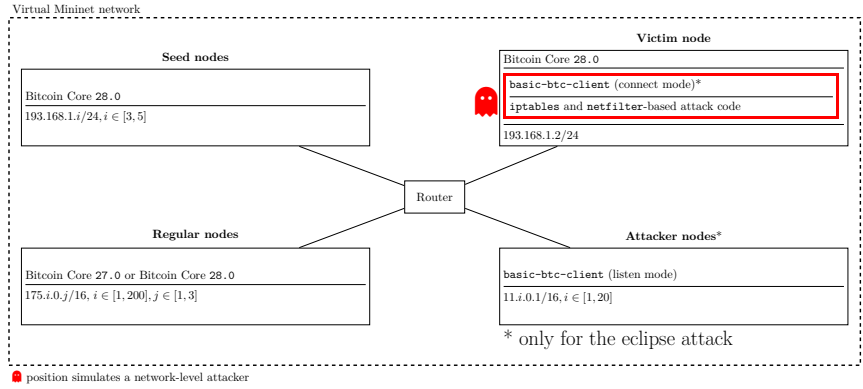

We built our own testbed, \testbed, for the purpose of evaluating the attack in
a controlled environment~(see \cref{sec:ethics} for ethical considerations).
It is based on the Mininet project~\cite{lantz2010network} and uses Docker
containers as hosts to build a virtual network topology.
We refer to \cref{subsec:testbed} for further information on \testbed.
We deployed the topology illustrated in \cref{fig:topology} on an Ubuntu
$24.04$~LTS server with $16$~cores and $94$~GiB RAM.
In addition to seed nodes, the network consisted of $621$ hosts and was
configured as follows:
\begin{itemize}
    \item $3$ hosts each in $200$ subnets, running either \btcdefaultversion or
        \btcversion;
    \item $1$ Debian Linux container with our implementation of the attack
        attached to a \netfilter queue as well as \btcversion~(the victim node).
        The attack code, which we implemented using the \netfilter framework and
        describe in greater detail in~\cref{subsec:impl_eclipse}, replays
        \code{PING} messages and uses \btcclient\footnote{\btcclient is our
            implementation of a minimal Bitcoin client
        (cf.~\cref{subsec:impl_eclipse} for a description).} to occupy the
        victim's inbound slots and announce adversarial addresses.
        An \code{iptables} rule set directs all packets on \gls{TCP} port
        \defaultport to the \netfilter queue.
        The victim and the attacker were placed in the same container due to
        implementation simplicity.
        The setup is representative of an in-path attacker as traffic must flow
        through them, \ie the adversary's position is different, but the traffic
        they can intercept and their capabilities remain the same.
        Separating the two components in our testbed is not expected to have any
        impact on the attack's evaluation;
    \item $20$ subnets containing $1$ host each running \btcclient and waiting
        for connections.
        The adversary uses these hosts to fill the victim's outgoing connection
        slots; and
    \item a block generation time of ten minutes.
\end{itemize}
All Bitcoin Core instances were configured with the default settings.
The only exception was the victim as we set the maximum number of connections to
$50$ which only has an impact on the number of incoming connections the victim
will allow, \ie $10$ outgoing, $1$ feeler and $39$ incoming connection slots.
As inbound slots are easier to monopolise, this setting has little to no impact
on the validity of our experiments.
The attack code was launched $24$~hours after the network was deployed in order
to allow the victim to arrive at a \enquote{stable} state with respect to
its peers and address tables.
$100\%$ of the victim node's outgoing and $\approx80\%$ of its incoming
connection slots were occupied when the attack started (see
\cref{fig:target_peer_count}
in
the appendix for the absolute number of connections).
This is comparable to what we observed on our measurement node and confirms that
the attack was evaluated under realistic conditions with respect to the victim's
connections.
The presented results refer to data collected $30$~minutes prior
to the attack until the victim was eclipsed.

\begin{figure}
    \centering
    \begin{minipage}[t]{0.325\linewidth}
        \resizebox{1\linewidth}{!}{
            \includegraphics[]{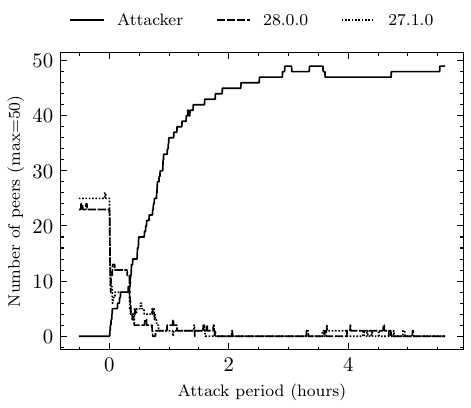}
        }
        \caption{The victim node's number of inbound and outbound peers by user
            agent before and during the attack.}
        \label{fig:target_user_agents}
    \end{minipage}%
    \hfill
    \begin{minipage}[t]{0.325\linewidth}
        \resizebox{1\linewidth}{!}{
            \includegraphics[]{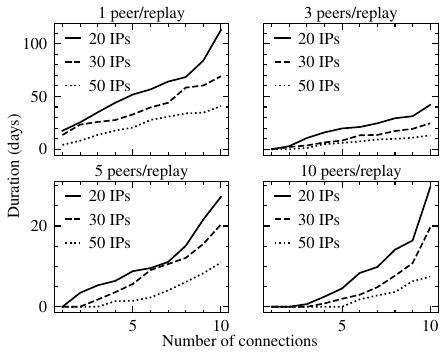}
        }
        \caption{Time required for the attacker to occupy a $30$-day old
            node's outgoing connection slots. %
        }
        \label{fig:addr_sim}
    \end{minipage}
    \hfill
    \begin{minipage}[t]{0.325\linewidth}
        \centering
        \resizebox{1\linewidth}{!}{
            \includegraphics{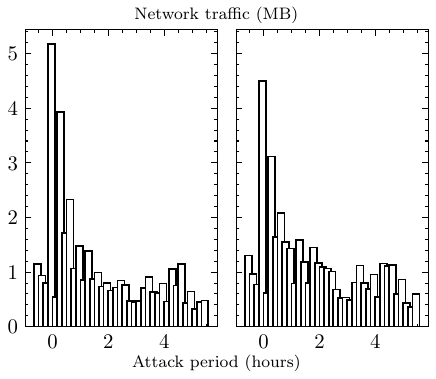}
        }
        \caption{Data sent~(left) and received~(right) by the victim's Bitcoin Core
        client before and during the attack in megabytes.}
        \label{fig:target_traffic}
    \end{minipage}
\end{figure}

\subsection{Feasibility}

\subparagraph{Message classification}

The attacker's ability to identify application messages in the \gls{TCP}
payloads is of significant value to the attack's success, especially with regard
to terminating connections efficiently.
In order to quantify how well the adversary is able to identify messages types
in real time, we compared the ground-truth messages logged by the victim's
Bitcoin Core client and the attack code's logs during the attack.
We examine only outgoing \code{PING}s because they are relevant for the attack.
We calculated commonly used classification metrics for these messages which
reveal a precision of $0.99$ and recall of $0.52$.
The high precision means that payloads classified as \code{PING}s were almost
always \code{PING}s.
The lower recall means that the program failed to detect approximately half of
the \code{PING}s sent by the victim node.
The main reason for that is that a \gls{TCP} segment may carry more than one
application message which inevitably leads to inaccuracies when only using
payload size as an indicator.

\subparagraph{Attack success}

\cref{fig:target_user_agents}
shows the victim node's number of connections differentiated by user agent, \ie
the client version advertised by a node.
The onset of the attack is clearly visible in two ways.
Firstly, there is a steep drop in the number of connections.
This initial decline is expected because the node had multiple connections which
were terminated by the attacker in quick succession.
Secondly, the number of peers advertising their user agent as \enquote{Attacker}
starts to rise.
However, the rise is not as steep as the decline of legitimate user agents.
This is because some of the closed connections are outgoing, and
the adversary has no control over when the victim will connect to them.
This is why the number of legitimate user agents sometimes increases
during the attack.
The victim is eclipsed after approximately five hours at the point when $49$ of
its connections are to the attacker.
The $50^{th}$ connection slot is only ever occupied temporarily when the victim
node initiates a feeler connection.

While these results demonstrate that the presented attack can be mounted
successfully, they do not show that, in reality, the attack requires more time.
This is because a target's address database will contain more entries.
In order to estimate how long it would take under more realistic conditions,
we simulated a simplified version of the selection of \emph{outgoing}
connections~(see \cref{subsec:impl_addrsim} for implementation details).
Incoming connections are not decisive for the duration of the attack as the
adversary initiates them.
Bitcoin Core sends \code{PING}s to peers every two minutes meaning that the
adversary can replay a packet on a given connection after four minutes.
The simulation assumes that the adversary replays a packet between the victim
node and a given peer every four minutes, \ie a new peer is chosen every
four minutes.
We performed the simulations using the address database of a $30$-day old
node, a varying number of addresses, and a varying number of concurrent
disconnections per minute.
The latter refers to the number of connections the adversary can close per
minute, \eg when set to one, the attacker closes a single connection per minute.

The estimated time required to occupy all ten outgoing connection slots with
adversarial addresses is shown in \cref{fig:addr_sim} for various combinations of
these parameters.
The worst case is if the adversary can only close a single connection at a time
and uses only $20$ different IP addresses.
In this case, they require around $100$ days to eclipse the victim node.
Using the same number of addresses, the required time is reduced to less than
$30$ days if five connections are closed at the same time.
In the best case, the adversary uses $50$ addresses and is able to always close
all ten connections at a go.
The attack then takes just $10$ days to succeed.

Aside from the fact that comparable attacks are no longer practical, our attack is
less expensive (w.r.t. the necessary network addresses) and faster, \eg the EREBUS attack takes five to six weeks.
This is because our attack relies more on exhausting entries in the victim's
address database instead of filling the entire database.
The best results are obtained by closing multiple connections at the
same time and having many addresses available.
The duration is shortened further if they can close multiple connections in
quick succession, which depends on how fast they can identify payloads
containing \code{PING} messages and/or how offensive they are willing to be.
The adversary can therefore assign higher precedence to stealth, lower resources
or speed making the attack more versatile than similar ones.

\subsection{Latency overhead}
\label{subsec:latency_eclipse}

We measured the latency added to each packet by our implementation during the
attack.
We recorded the time from the program first accessing a packet in the \netfilter
queue to a decision being made and the packet being returned to the kernel.
The data was collected from the onset of the attack until its completion.

The results suggest that our implementation is quite efficient and issues a
verdict on a packet within a median time of $\micros{32.6}$ and mean time
of $\micros{39}$ (see~\cref{fig:delay_eclipse} the appendix for a plot).
That equates to a mean throughput of $\approx25,641pps$ regardless of packet
size.
in
We conclude that the delays induced by the extra filtering layer are negligible
and do not have a negative impact on the feasibility of the attack.
We argue that delays in the range of tens of microseconds are likely to go
unnoticed by operators.
Besides, the attack is implemented in user-space code, and would likely be more
efficient if it made use of lower-level functions.

\subsection{Attack footprint}

The \gls{TCP} connection is closed by the remote side regularly via the four-way
handshake which does not suggest any malicious interference.
The victim node's Bitcoin Core client writes an \code{ECONNRESET}
error~\cite{linuxerrors} to the log file which does not indicate any malicious
activity either.
Among other reasons, such an error may be issued by the kernel when the client
attempts to write to a connection the remote peer considers to be closed as is
the case here.
We therefore examine whether the attack leaves any evidence at the network
layer.
\cref{fig:target_traffic}
shows the total network traffic recorded by the victim's Bitcoin Core client
prior to and during the attack.
It shows an unmistakable peak in both sent and received traffic when the attack
starts.
This can be explained by the fact that many connections are closed and opened in
quick succession at that point.
As the number of connections to the attacker increases, the rate of traffic
on average reduces.
This is because \btcclient nodes do not initiate any unsolicited communication.
However, the traffic pattern changes in that it is burstier compared to
before the attack started.
The reasons for the bursts are either the opening/closing of connections or the
victim node transmitting application messages.

The main indicator of the attack is likely to be the sudden onset of closed
connections in quick succession which can be countered by staggering the attack,
\eg attack at most a certain number of connections per time interval.
However, as evidenced by the simulation results in \cref{fig:addr_sim}, the
attack is significantly faster when connections are closed speedily.
The adversary therefore needs to assign higher precedence to either speed or
stealth.
The change in traffic patterns might also make an attentive node operator
suspicious leading them to dig deeper and discover that all of their incoming
connections are from one IP address.
This risk can be mitigated easily by imitating Bitcoin Core's behaviour in
\btcclient.
However, none of this actually provides insight on to why connections are being
closed.

\section{Evaluation: Downgrade attack}
\label{sec:eval_downgrade}

In the following, we describe our evaluation setup and then present results
concerning various aspects of the downgrade attack.
Multiple results and graphs are omitted due to space constraints but are
provided in \cref{fig:supp_downgrade} as part of the appendix.

\subsection{Methodology}

We used the network shown in \cref{fig:topology}
(without attacker nodes) and configured the victim
container to run the user-space implementation of the attack attached to a \netfilter
queue as well as a \btcversion client.
An \code{iptables} rule set is used to direct all \gls{TCP} packets on
port \defaultport from the client to the \netfilter queue.
The attack code, detailed in~\cref{subsec:impl_downgrade}, monitors the victim's network traffic and forges \code{RST} packets by
modifying a legitimate packet when it detects a new \gls{V2} connection attempt.
Similar to the eclipse attack, the attack code was launched $24$~hours after the
network was deployed in order to allow the victim node to arrive at a
\enquote{stable} state.
The entire network was then restarted at the onset of the attack so as to model
basic churn as connections in the private network (and in the public network)
are otherwise very long-lived.
By doing so, we ensure that the victim node receives incoming connections while
the attack code is attached.
While this does not represent natural churn, the best time to perform the
downgrade attack is after a reboot as all connections can be downgraded
simultaneously, making legacy attacks possible sooner.
The results presented in what follows refer to data collected $30$~minutes prior
to launching the attack and during the first two hours.

\subsection{Feasibility}

\begin{table}
    \centering
    \caption{
        Excerpt of a packet trace between a node (\code{175.6.0.2}) and the victim
        node (\code{193.168.1.2}) in the testbed as captured and displayed by
        Wireshark.
        Plain \code{ACK} packets have been omitted for brevity.
        The attacker forges a \code{RST} packet (\#9) from the responder
        during the key establishment phase.
        The initiator reacts by reconnecting to the victim using \gls{V1}.
    }
    \label{tab:pkt_trace}
    \scriptsize
    \catcode`"=9
    \begin{tabularx}{\linewidth}{p{0.02\linewidth}llllp{0.025\linewidth}X}
    \toprule
    \textbf{No.} & \textbf{Time} & \textbf{Src} & \textbf{Dst} &
    \textbf{Prot.} & \textbf{Len.} & \textbf{Info} \\
    \midrule

    \begin{filecontents*}{./data/downgrade-victim-responder-relative.csv}
"No.","Time","Source","Destination","Protocol","Length","TcpLength","Info","Destination Port","Source Port"
"1","09:37:56.858152","175.6.0.2","193.168.1.2","TCP","76","0","44120$>$8333 [SYN] Seq=0","8333","44120"
"2","09:37:56.858284","193.168.1.2","175.6.0.2","TCP","76","0","8333$>$44120 [SYN;ACK] Seq=0 Ack=1","44120","8333"
"3","09:37:56.864617","175.6.0.2","193.168.1.2","TCP","68","0","44120$>$8333 [ACK] Seq=1 Ack=1","8333","44120"
"4","09:37:56.889652","175.6.0.2","193.168.1.2","TCP","2964","2896","44120$>$8333 [PSH;ACK] Seq=1 Ack=1","8333","44120"
"5","09:37:56.889674","175.6.0.2","193.168.1.2","TCP","693","625","44120$>$8333 [PSH;ACK] Seq=2897 Ack=1","8333","44120"
\textbf{"9"},\textbf{"09:37:56.890003"},\textbf{"193.168.1.2"},\textbf{"175.6.0.2"},\textbf{"TCP"},\textbf{"68"},\textbf{"0"},\textbf{"8333$>$44120 [RST;ACK] Seq=1 Ack=3522"},"44120","8333"
"10","09:37:56.979774","175.6.0.2","193.168.1.2","TCP","76","0","44126$>$8333 [SYN] Seq=0","8333","44126"
"11","09:37:56.979889","193.168.1.2","175.6.0.2","TCP","76","0","8333$>$44126 [SYN;ACK] Seq=0 Ack=1","44126","8333"
"12","09:37:56.985895","175.6.0.2","193.168.1.2","TCP","68","0","44126$>$8333 [ACK] Seq=1 Ack=1","8333","44126"
"13","09:37:57.006229","175.6.0.2","193.168.1.2","Bitcoin","194","126","version","8333","44126"
"15","09:37:57.006425","193.168.1.2","175.6.0.2","Bitcoin","194","126","version","44126","8333"

\end{filecontents*}
\csvreader[
        late after line=\\,
    ]{./data/downgrade-victim-responder-relative.csv}
    {1=\No,2=\Time,3=\Source,4=\Destination,5=\Protocol,6=\Length,7=\TcpLength,8=\Info,9=\Dport}
    {\ttfamily\No & \ttfamily\Time & \ttfamily\Source & \ttfamily\Destination & \ttfamily\Protocol & \ttfamily\TcpLength & \ttfamily\Info}

    \bottomrule
    \end{tabularx}
\end{table}

We first demonstrate that the attack can indeed be carried out by a
network-level adversary.
We captured the network traffic at the victim node and show an excerpt of the
packet trace in~\cref{tab:pkt_trace}.
It shows a connection
establishment between a regular node~($175.6.0.2$) and the victim
node~($193.168.1.2$) while the attack code was running.
Packets $\#4$ to $\#8$ show the initial parts of the key exchange phase.
The code does not tamper with these packets but we now know that the
connection is being setup for \gls{V2}.
Therefore, the program changes packet $\#9$ to a \code{RST} from the responder.
In response, the initiator opens a new \gls{TCP} connection~($\#10$ to $\#12$)
and then starts the unencrypted handshake by sending a \code{VERSION}
immediately afterwards.
The connection has been successfully downgraded at this point.

\subsection{Attack success}

\begin{figure}
    \centering
    \includegraphics[height=4.5cm]{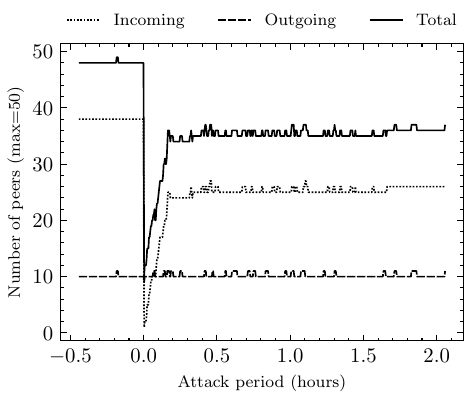}
    \caption{The victim node's number of peers before and during the attack.
        The connections established after the restart were all using \gls{V1}
        unlike before the attack.
    }
    \label{fig:target_transport}
\end{figure}

We now show that the attack was successful in that all of a victim's
connections were downgraded to the \gls{V1} protocol.
$100\%$ of the victim node's connection slots were occupied prior to the attack
and were using \gls{V2} as expected.
\cref{fig:target_transport}
shows
the number of peers the victim node maintained prior to and during the attack.
The onset of the attack is clearly visible at the point when the node's
connection count drops to zero.
The figure shows that the victim node quickly established all of its outgoing
connections, while it took more time before its incoming connection slots were
filled.
This is consistent with previous work~\cite{tran2020stealthier} and
our measurements in the mainnet.
However, $100\%$ of the connections that were established during the attack used
\gls{V1}.

In reality, the attack requires more time if the victim node is already
participating in the \gls{P2P} network.
The adversary may need to wait for a reboot of the victim~(cf.~\cref{subsec:downgrade_reboot}).
However, depending on the adversary's ultimate objective, a certain
percentage of V1 peers may be sufficient.
In that case, the attack duration is shorter and may not require a reboot.
In fact, our measurements showed that at no point were more than
$50\%$ of the outgoing connections and $70\%$ of the incoming connections using
\gls{V2}.
This suggests that the attack is likely less time-consuming in practice.

\subsection{Latency overhead}

We measured the latency added by our implementation of the attack during the
first two hours.
Specifically, we recorded the duration from the program first reading a packet
from the \netfilter queue to a decision being made and it being returned to the
kernel.

The results are similar to the implementation of the eclipse attack in
~\cref{subsec:latency_eclipse}.
The program issues a verdict on a packet within a median time of $\micros{41}$
and mean time of $\micros{49}$.
This leads to a mean throughput of $\approx20,408pps$ regardless of packet size.
This implementation can also be made more efficient by making use of lower-level
functions.
We refer to \cref{fig:delay_downgrade}
in the appendix
for a plot of these results.

\subsection{Attack footprint}

Although the window of opportunity for this attack is long enough for an on-path
adversary to inject a \code{RST} packet into the stream, we chose to implement
it as an in-path attack.
This is because on-path attack systems can generally be detected by observing
anomalies in the traffic flows~\cite{marczak2015analysis, weaver2009detecting}.
One would simply notice the arrival of a \code{RST} packet despite already
receiving the legitimate key exchange packets.
The main reason for concern might be the fact that none of a node's
connections, including connections the victim node had prior to the attack, are
using \gls{V2}.
However, the lack of derivations from the protocol specification would make it
difficult to identify the attack without out-of-band communication.

\section{Countermeasures}
\label{sec:countermeasures}

Our analysis of \gls{V2} uncovered conceptual weaknesses that we exploit for the
attacks this work.
Completely mitigating the attacks may not be possible: closing connections on
failed decryption appears to be standard practice and a full protocol
negotiation mechanism requires updates to past releases.
However, there are certain features of Bitcoin and \gls{V2} that are conducive
for attacks.
In the following, we pinpoint these problems (\textbf{P}), propose possible
countermeasures~(\textbf{C}) for each and evaluate some short-term measures.

\subsection{Primary countermeasures}

This first set of measures relates directly to making network-level attacks less
viable in general.
We believe that these measures are effective on their own, but may require
significant protocol changes and additional research to fully understand
their expected impact on Bitcoin.

\newissue[ident]{%
    Network-level information leakage%
}
{
    Classification at the network level without the need
    for any state information is reasonably accurate.
    We can expect even higher accuracy if the adversary uses sophisticated
    fingerprinting techniques, \eg leveraging that Bitcoin is mainly a
    request-response protocol.
}

\newmeasure[]{
    Bitcoin needs to obfuscate \gls{TCP} payloads' features by including a
    length-hiding scheme.
    Examining what an appropriate mechanism for Bitcoin would be exceeds the
    scope of this work, but we refer to a recent work~\cite{ndolo2024payment}
    for a comprehensive discussion on padding strategies for the
    \glsentrylong{LN}.
    The main points apply to Bitcoin.
}

\newissue[]{%
    Default port%
}{
    \gls{TCP} port \defaultport is the
    default for \gls{P2P} communication and was used by over $95\%$ of the
    reachable nodes according to the Bitnodes crawler at the time of writing.
    As \gls{V2} is not self-revealing, a default port is somewhat
    self-defeating.
}
\newmeasure[port]{
    Using a non-default port forces the adversary to monitor \emph{all}
    network traffic and not just on port \defaultport.
    Transitioning to non-default ports has been raised
    previously~\cite{apostolaki2017hijacking, bitcoincoreports} but we are not
    aware of any related changes.
    While we generally support this recommendation, we argue that it is only
    effective in combination with traffic shaping.
    The new port can otherwise possibly still be inferred from traffic
    patterns~(cf. \textbf{P\ref{issue:ident}}).
    We also acknowledge that random ports would complicate matters for network
    devices such as firewalls.
}

\newissue[]{%
    Network-level protocol negotiation%
}{
    \relevantbip encourages
    clients to retry with the V1 protocol if the connection is
    closed immediately after the \gls{TCP} handshake.
    This makes it possible for a network-level adversary to downgrade a
    connection.
}

\newsubmeasure[
    Protocol version negotiation should be handled by the application layer as
    is commonly done.
    For instance, HTTP's upgrade mechanism over TLS could be used to negotiate
    the protocol after connection establishment as is done in
    Ripple~\cite{ripple}.
    However, any such approach would only be meaningful in combination with
    authentication~(cf. \textbf{P\ref{issue:auth}}).
    ]{
    Currently, feeler connections are closed immediately after the application
    handshake.
    This means that a node not only discovers whether an address is reachable,
    but also which transport version the remote node supports.
    When the address is selected as an outgoing peer, the local node should
    attempt to only connect with the discovered protocol.
}{
    A further measure is to discontinue this behaviour when the adoption
    of \gls{V2} is considered to be mature enough or after a certain number of
    releases past when \gls{V2} was activated.
    Older clients would still be able to connect to the network as they, as the
    initiators, decide whether to perform the key exchange or not.
}

\newissue[auth]{%
    Lack of authentication%
}{
    Our analysis shows that a \gls{P2P} connection can be downgraded upon
    reconnection even though the peers had been previously using \gls{V2}.
}

\newmeasure[]{
    We suggest adding transport protocol information to the state maintained on
    peers and only reconnecting using the stored version or higher.
    This is especially relevant for connections whose main purpose is to protect
    a node from attacks.
    If \textbf{C\ref{ctr:port}} were to be adopted, authentication becomes
    indispensable in order to ensure the legitimacy of the remote peer.
    In the case of \textbf{C3B}, feeler connections should include
    authentication.
}

\subsection{Secondary countermeasures}

The next set of measures addresses additional issues we identified in the course
of this work.
They, however, cannot secure the \gls{P2P} overlay from network attacks on their
own.

\newissue[]{%
    Unrestricted incoming connections%
}{
    The resources required for attacks are lowered significantly as all incoming
    connections can be from one IP address.
    In fact, measurement data shows that at
    least $75\%$ of incoming connections were from distinct $/16$ prefix
    groups~(cf.~\cref{fig:peer_prefixes} in the appendix).
}

\newmeasure[incoming]{
    We recommend limiting the number of incoming connections per
    network prefix or at least per IP address to raise adversarial costs.
    As incoming slots are considered scarce~\cite{grundmann2022on}, the
    constraints need not be as strict as for outbound connections
    and should only came into effect when most slots are occupied to mitigate
    partition risks.
}

\newissue[]{%
    Unverified service flags%
}{
    Service flags are advertisements made by nodes to announce protocol features
    they support.
    They are part of \code{ADDR} messages and can be spoofed or falsified
    trivially.
    Nodes can therefore simply announce the favourable features without actually
    supporting them, much to the benefit of an adversary.
}

\newsubmeasure[]{
    Service flags, at least those required for outbound connections, should be
    validated, \eg using a challenge-response protocol.
    A node that advertises the \code{NETWORK} flag should be able to respond
    with a valid \code{BLOCK} when queried for a \emph{random} block
    hash.
    An adversary can still fake their service flags, but they would be forced to
    maintain protocol state.
}{
    Although less preferable, service flags should not be taken into
    consideration when selecting outbound peers.
    As it stands, they do not do not provide any real protection.
}

\newissue[]{%
    Bypassable synchronisation check%
}{
    Among others, a node checks that an outgoing peer is synchronised to the same
    blockchain height as they are.
    If not, the node sends a \code{GETHEADERS} message asking for recent block
    hashes and closes the connection if it does not receive a valid response
    within a timeout.
    The sustainability of eclipse attacks is enhanced by the fact that
    an adversary can simply send the exact same \code{GETHEADERS} message
    to the node, wait for their \code{HEADERS} reply and then send the
    response back.
}

\newmeasure[]{
    Nodes should not respond to \code{GETHEADERS} from peers that are up for
    possible disconnection.
    This would force the adversary to maintain state via legitimate connections.
}

\newissue[]{%
    Lack of anomaly detection%
}{
    Bitcoin Core does not include any monitoring or anomaly detection.
    This makes it very hard for node operators to detect attacks in time.
}

\newmeasure[]{
    Anomaly detection mechanisms should be deployed to support node operators in
    securing their nodes.
    It should issue warnings and alerts about
    unusual events such as unusually frequent disconnections or extended periods
    without receiving any data.
    Monitoring would not mitigate the attacks but it would allow operators to react accordingly.
}

\subsection{Evaluation of short-term countermeasures}
\label{subsec:eval_countermeasures}

Although the primary measures are the most effective, they are impractical in
the near term as they require protocol changes and further research.
The secondary measures, on the other hand, are insufficient on their own but
several can already be deployed immediately.
To assess their practical viability, we implemented \textbf{C3C} and
\textbf{C\ref{ctr:incoming}} in \btcversion and ran two nodes with these changes
on the public network.
Alongside \textbf{C3C}, which prevents any connection attempt from being retried
following a \gls{TCP} reset, we implemented \textbf{C\ref{ctr:incoming}} such
that requests from prefix groups the node already has connections from are only
accepted when less than $x\%$~($x \in \{90, 95\}$) of the incoming slots are
full.
We monitored both nodes' connection statistics over $30$~days.

The percentage of occupied incoming slots and of outgoing \gls{V2} peers is shown in
~\cref{fig:conns_w_countermeasures} in the appendix.
It reveals variations in how fast nodes fill their incoming connection slots.
The $90\%$ threshold was met after $8$~days and the remaining slots were
gradually filled up to full capacity despite the restrictions on the connections
to accept.
One node received connection requests very slowly such that the $95\%$ threshold
had not yet been met at the time of writing.
The data also shows that both nodes had outgoing V1 connections despite
supporting \gls{V2} and \textbf{C3C}.
This is because Bitcoin Core uses the service flags propagated in \code{ADDR}
messages to determine whether to attempt a \gls{V2} connection.
Manual inspection of the relayed service flags after connection revealed
that several of our V1 peers did in fact support \gls{V2} but this had not
propagated to our nodes at the time of connection establishment.
Overall, the data indicates that all incoming slots will eventually be filled
despite \textbf{C\ref{ctr:incoming}} and the maximum number of outgoing
connections is also maintained under \textbf{C3C}.
We are therefore confident that these measures can be deployed immediately
without adversely affecting the \gls{P2P} network.

\section{Related Work}%
\label{sec:related}

Due to the magnitude of research on attacks in Bitcoin, we limit the following
discussion to network-level attacks and refer to surveys~\cite{bonneau2015sok,
franzoni2022sok} for extensive overviews on various attacks.

Eclipse attacks in Bitcoin were first discussed in
2015~\cite{heilman2015eclipse} at a time when Bitcoin had several shortcomings
in its management of nodes' addresses.
Despite the deployment of multiple countermeasures, multiple works later
presented eclipse
attacks~\cite{apostolaki2017hijacking,fan2021conman,tran2020stealthier}.
These attacks are closely related to our work as they share a common attacker
model and attack objective.
However, our eclipse attack differs from all of these attacks in multiple
aspects.
It does not involve hijacking network traffic and is designed to work in the
presence of encrypted communication.
Additionally, none of the attacks are still viable under \gls{V2} as originally
described.
Furthermore, our attack does not require a reboot of the target node unlike the
attacks presented in~\cite{heilman2015eclipse,tran2020stealthier}.
The attack in~\cite{heilman2015eclipse} differs from ours in more ways besides
not operating at the network layer.
Most notably, we do not exploit protocol vulnerabilities to initiate connections
from the victim to the attacker.
Unlike the earlier works, we also discusses sustaining the attack in detail.
Eclipse attacks have also been studied in related \gls{P2P} networks such as
Ethereum~\cite{henningsen2019eclipsing, marcus2018low},
Monero~\cite{shi2025eclipse} and \glsentryshort{IPFS}~\cite{pruenster2022total}.

Some of these related works~\cite{apostolaki2017hijacking, tran2020stealthier}
also performed measurements of the Bitcoin mainnet.
However, they do not cover the \gls{V2} protocol as they were conducted at a
time when \relevantbip was not yet finalised.
This work is, to the best of our knowledge, the first to study the
\gls{P2P} layer following the specification of \gls{V2}.
Existing work briefly describes \gls{V2} as a countermeasure to the presented
attack~\cite{zou2024unveiling}.

Bitcoin's \gls{P2P} network has long been subject to traffic analysis attacks
targeting privacy~%
\cite{biryukov2014deanonymisation,biryukov2015bitcoin,koshy2014analysis} and
topology inference~\cite{neudecker2016timing}.
\relevantbip aims to increase the cost of such attacks by encrypting
messages.
While recent works on traffic analysis focus on the Lightning
Network~\cite{ndolo2024payment, arx2023revelio}, the threat of traffic analysis
is not specific to blockchain networks~\cite{dyer2012peek, miller2014i,
raymond2000designing}.

\section{Conclusion}%
\label{sec:conclusion}

This paper presented the first study of Bitcoin's new \gls{V2} protocol.
In doing so, we contribute to the further understanding of the protocol and
offer an additional source of reference.
We studied \gls{V2}'s ability to secure the overlay from network-layer attacks.
While we conclude that \gls{V2} successfully hinders previously-known,
network-layer attacks, we present an alternative path to eclipse attacks.
By exploiting properties introduced by the usage of encrypted communication
channels, we developed an eclipse attack that is based on closing connections by
replaying \gls{TCP} payloads.
Our analysis of the attack showed that it is highly effective.
We then presented a downgrade attack that takes advantage of how compatibility
with the V1 protocol is achieved.
It reopens the door to attacks that are no longer possible in the presence of
encrypted communication.
In conclusion, we proposed a series of countermeasures that can be taken in
response to the findings in this work.
Some of them, \eg defining a length-hiding scheme, require additional research
and present interesting avenues for future work.

\bibliography{paper.bib}

\section*{Ethical Considerations}
\label{sec:ethics}

The primary goal of this work is to contribute to further advancement of the
security and resilience of the Bitcoin network for all involved users.
Uncovering, presenting, and addressing potential concerns in the network is a
central part of that process.
Prior to submission, we shared a copy of the manuscript and informed Bitcoin
Core's security team about our intention to publish our results.
We are yet to receive a response, but note that we are prepared to discuss the
suggested countermeasures and to cooperate on implementing measures in response
to this work.

As far as the practical evaluation of the attacks is concerned, we followed the
guidelines of the Menlo report~\cite{bailey2012the} and general security
research best practices.
In particular, with the exception of conducting measurements using our own node,
we did not interact with the public mainnet in any way.
The measurements neither store nor reveal any information about other nodes in
the network, and are based entirely on normal node operation.
All adversarial experiments were conducted in our private testbed~(in regtest
mode) and only consists of nodes set up for the precise purpose.

We will not publish our implementations of the attacks as we cannot control
how they are used.
We, however, made the code available to reviewers in entirety and will consider
doing the same to researchers upon request.

\section*{Open Science}
\label{sec:open_science}

\newcounter{item}

In line with the principles of open science, we have uploaded all of our
source code to an anonymised repository.
We have made our best efforts to document it clearly enough for independent use.
The following artefacts are available at the respective repository:
\begin{enumerate}
    \item our proof-of-concept implementation of both the eclipse and downgrade
        attacks is only available for reviewers;
    \item \btcclient which is a minimal Bitcoin client\footnote{\pocurl};

    \item the implementation of our Mininet-based testbed.\footnote{\testbedurl}
        It was designed with generality in mind and is potentially useful for
        similar research projects.
        The documentation contains details on how to reproduce our experiments;
        and

    \item our simulator for the selection of outgoing addresses.\footnote{\simurl}
        The repository includes our node's dump of its address tables as
        obtained from Bitcoin Core using the \code{getrawaddrman} RPC.
\end{enumerate}

Subsequent to peer review, the artefacts will be published on GitHub for
general access.
The only exception is our implementation of the attacks which we withhold to
mitigate potential misuse.
We consider the paper detailed enough for readers to reproduce with their own
implementation, if necessary.
We will, however, provide the code for research purposes only upon request
contingent on it not being distributed beyond the intended scope.

\appendix

\section{Measurement Study of \gls{V2} in the Public Network}
\label[appendix]{sec:measurements}

In an effort to understand the extent and impact of \gls{V2} on the Bitcoin
network, we conducted measurements in the mainnet.
The results shown in this work refer to data collected between
\measurementsstart and \measurementsend.
The measurements are performed passively using our own freshly-deployed node
running \btcversion in the mainnet.
The node is reachable via a public IP address, accepts incoming connections via
\tcpip and is configured with the default settings.
To the best of our knowledge, this is the first empirical
study of \gls{V2} in the public network.

\subparagraph{Number of peers}

\begin{figure*}
    \begin{minipage}[t]{0.5\linewidth}\centering
    \includegraphics[height=4.5cm]{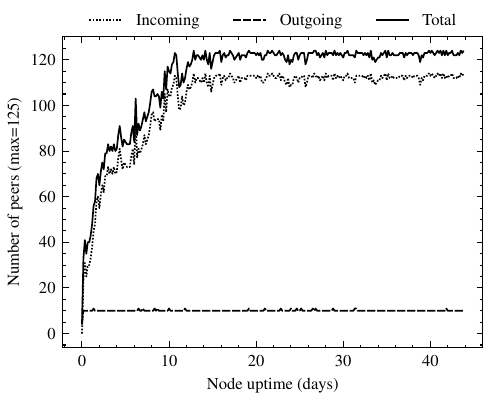}
    \caption{Number of inbound and outbound peers.}
    \label{fig:peer_count}
    \end{minipage}
    \hfill
    \begin{minipage}[t]{0.5\linewidth}\centering
    \includegraphics[height=4.5cm]{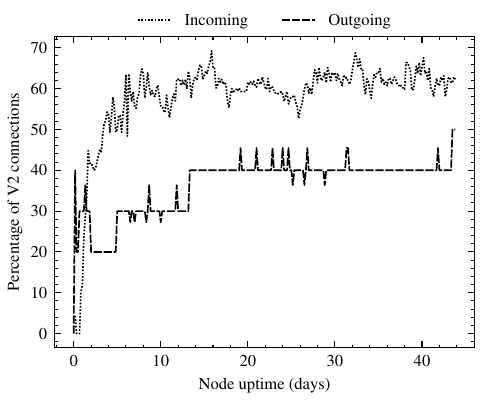}
    \caption{Percentage of V2 connections maintained by the measurement node
    distinguished by connection type.}
    \label{fig:peer_transport}
    \end{minipage}%
    \newline
    \begin{minipage}[t]{0.5\linewidth}\centering
    \includegraphics[height=4.5cm]{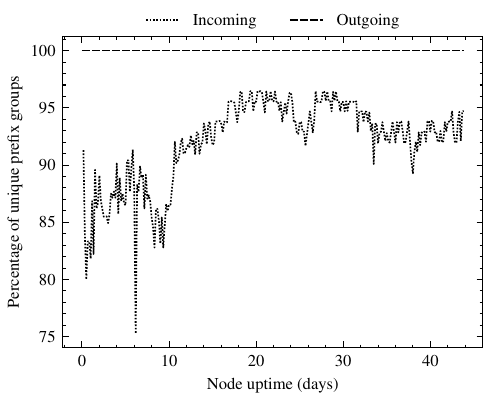}
    \caption{Percentage of peers from distinct $/16$ prefix groups.}
    \label{fig:peer_prefixes}
    \end{minipage}
    \hfill
    \begin{minipage}[t]{0.5\linewidth}\centering
    \includegraphics[height=4.5cm]{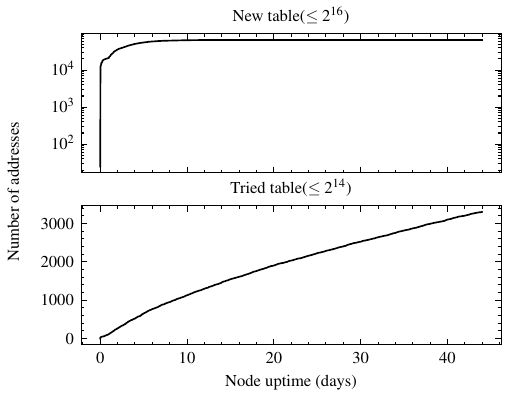}
    \caption{Number of entries in the measurement node's address
    database.}
    \label{fig:addrs}
    \end{minipage}%
\end{figure*}

The first set of results provides general data on the measurement node.
\cref{fig:peer_count}
shows
its number of connections based on snapshots taken every four hours.
The figure shows that the node established the maximum of ten outgoing
connections within a short period of time and steadily maintained that many
connections.
Spikes in the number of outgoing connections are due to \emph{feeler} connections
that test the reachability of stored addresses.
On the other hand, the number of incoming connections gradually increases over
time such that over $90\%$ of its available connection slots were filled after
$14$ days.
Both of these observations are consistent with reports in previous
works~\cite{tran2020stealthier}.

\subparagraph{Percentage of \gls{V2} peers}

\begin{figure}
    \centering
    \includegraphics[scale=0.9]{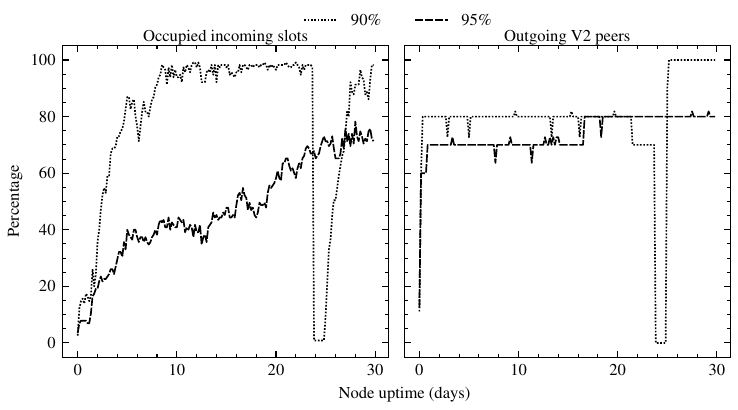}
    \caption{Percentage of occupied incoming slots and of outbound \gls{V2}
        peers after activating \textbf{C\ref{ctr:incoming}} with $\{90\%, 95\%\}$
        and \textbf{C3C}.
        The sudden drop on days $24$ and $25$ is due to an unexpected shutdown
        of the host running the affected measurement node.
    }
    \label{fig:conns_w_countermeasures}
\end{figure}

\cref{fig:peer_transport}
shows
the share of connections using the \gls{V2} protocol.
It shows that at most $50\%$ of the node's outgoing connections are \gls{V2}
connections.
This is somewhat unexpected as roughly $60\%$\footnote{According to data
gathered by the Bitnodes crawler: \url{https://bitnodes.io}.} of the reachable
nodes in the mainnet were using \btcdefaultversion and above as of
\measurementsstart.
As the connections tend to be long-lived, the percentage of \gls{V2} peers does
not fluctuate much.
This seems to suggest that potential outgoing peers should be selected with even
more care if support for encrypted communication is of importance as such
connections are likely there to stay.
In the case of incoming connections, the number slowly increases as more nodes
initiate connections to the measurement node.

\subparagraph{Percentage of unique prefix groups}

\cref{fig:peer_prefixes} shows the share of distinct $/16$ prefix groups
connected to the measurement node.
It shows that $100\%$ of its outgoing connections were to different prefix
groups.
This is expected as Bitcoin Core only initiates connections to prefix
groups that are not already part of the node's outgoing connections
to increase the cost of eclipse attacks.
Although no such restrictions exist on incoming connections,
over $75\%$ of them consistently originated from distinct prefix groups.

\subparagraph{Address tables}

\cref{fig:addrs}
shows the progression of the number of entries in the respective address
management tables during the measurement period.
We recorded the size of the measurement node's new and tried tables whenever an
entry was added to either table.
The figure shows that new table was already almost at full capacity
after approximately five days in the network.
On the other hand, the tried table was at roughly $20\%$ of its capacity at the
end of the measurement period.

\section{\gls{V2} message classification}
\label[appendix]{app:classification}

As described in \cref{subsec:classification},
the information available in the \gls{TCP} header suffices to weaken
confidentiality against passive attacks.
For illustration, we show how to identify message types using
the \code{INV} message.
It is used to advertise objects, \eg transactions or blocks, and features the
\code{INV\_VEC} data structure.

\begin{figure}
    \centering
    \begin{subfigure}[b]{0.99\textwidth}
    \centering%
        \begin{tikzpicture}
            \tikzstyle{label}=[above,midway,sloped,]

            \node[draw, rectangle] (varint) [] {
                \begin{tabular}{l|l}
                    Size (B) & Description \\
                    \midrule
                    1 & $[0, 253)$\\
                    3 & $[254, 2^{16}-1]  $\\
                    5 & $[2^{16}, 2^{32}-1]$\\
                    9 & $\geq 2^{32}$\\
                \end{tabular}
            };
        \end{tikzpicture}
    \caption{
        Size of the variable length integer encoding used for the
        length of application messages.
    }
    \label{fig:varint}
    \end{subfigure}
    \hfill
    \begin{subfigure}[b]{0.99\textwidth}
    \label{fig:inv}
    \centering%
        \begin{tikzpicture}
            \tikzstyle{label}=[above,midway,sloped,]
            \node[draw, rectangle] (payload) {$20B$ + $1B$ + \code{INV}};
            \node[draw, rectangle,anchor=north west,yshift=-25pt]
              (inv) at (payload.south west) {
                \begin{tabular}{l|l}
                    Size (B) & Desc. \\
                    \midrule
                    1+ & count \\
                    36$\times$0+ & inventory \\
                \end{tabular}
            };
            \node[draw, rectangle,xshift=0pt] (invvec) [right=of inv]{
                \begin{tabular}{l|l}
                    Size (B) & Desc. \\
                    \midrule
                    4 & type \\
                    32 & hash \\
                \end{tabular}
            };
            \draw[-] (payload) -- (inv.north -| payload);
            \draw[-] (inv) -- (invvec);
        \end{tikzpicture}
        \caption{
        Structure and size of a \gls{V2} \gls{TCP} payload transporting an
        \code{INV} message.
        It contains the mandatory fields (length descriptor, authentication tag
        and flags), one byte encoding the message type and the \code{INV} message.
        The \code{INV} message contains a variable number of inventory vectors
        which is encoded using the scheme in \cref{fig:varint}.
    }
    \end{subfigure}
    \vfill
    \begin{subfigure}[t]{0.99\textwidth}
        \centering%
        \begin{tikzpicture}
            \tikzstyle{label}=[above,midway,sloped,]
            \node[draw, rectangle] (payload) {$20B$ + $1B$ + \code{ADDR}};
            \node[draw, rectangle,anchor=north west,yshift=-25pt]
              (inv) at (payload.south west) {
                \begin{tabular}{l|l}
                    Size (B) & Desc. \\
                    \midrule
                    1+ & count \\
                    30$\times$0+ & address list \\
                \end{tabular}
            };
            \node[draw,rectangle, anchor=south west,xshift=15pt]
              (invvec) at (inv.south east) {
              \begin{tabular}{l|l}
                    Size (B) & Desc. \\
                    \midrule
                    4  & timestamp \\
                    8  & services \\
                    16 & IP address \\
                    2  & port number \\
                \end{tabular}
            };
            \draw[-] (payload) -- (inv.north -| payload);
            \draw[-] (inv) -- (inv -| invvec.west);
        \end{tikzpicture}
        \caption{
        Structure and size of a \gls{V2} \gls{TCP} payload transporting an
        \code{ADDR} message.
        It contains the mandatory fields (length descriptor, authentication tag
        and flags), one byte encoding the message type and the \code{ADDR} message.
        The \code{ADDR} message contains a variable number of address lists
        which is encoded using the scheme in \cref{fig:varint}.
    }
    \label{fig:addr}
    \end{subfigure}
    \vfill
    \begin{subfigure}[t]{0.99\textwidth}
        \centering%
        \begin{tikzpicture}
            \tikzstyle{label}=[above,midway,sloped,]
            \node[draw, rectangle] (payload) {$20B$ + $1B$ + \code{HEADERS}};
            \node[draw, rectangle,anchor=north west,yshift=-25pt]
              (inv) at (payload.south west) {
                \begin{tabular}{l|l}
                    Size (B) & Desc. \\
                    \midrule
                    1+ & count \\
                    81$\times$0+ & headers \\
                \end{tabular}
            };
            \node[draw,rectangle, anchor=south west,xshift=15pt]
              (invvec) at (inv.south east) {
                \begin{tabular}{l|l}
                    Size (B) & Desc. \\
                    \midrule
                    4 & version \\
                    32 & prev. block \\
                    32 & merkle root \\
                     4 & timestamp \\
                     4 & difficulty \\
                     4 & nonce \\
                     1 & 0 \\
                \end{tabular}
            };
            \draw[-] (payload) -- (inv.north -| payload);
            \draw[-] (inv) -- (inv -| invvec.west);
        \end{tikzpicture}
        \caption{
        Structure and size of a \gls{V2} \gls{TCP} payload transporting a
        \code{HEADERS} message.
        It contains the mandatory fields (length descriptor, authentication tag
        and flags), one byte encoding the message type and the \code{HEADERS} message.
        The \code{HEADERS} message contains a variable number of headers
        which is encoded using the scheme in \cref{fig:varint}.
    }
        \label{fig:headers}
    \end{subfigure}
    \caption{The format and size (in bytes) of select application messages as
    defined in the protocol.}
    \label{fig:proto_msgs}
\end{figure}
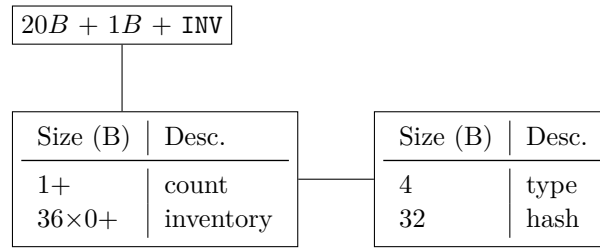
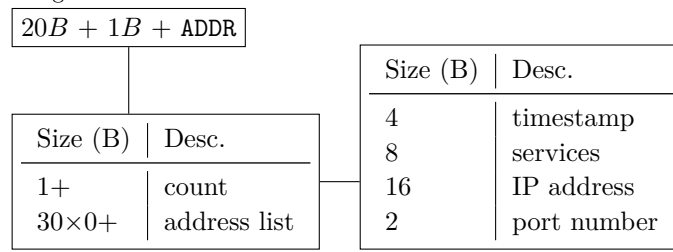
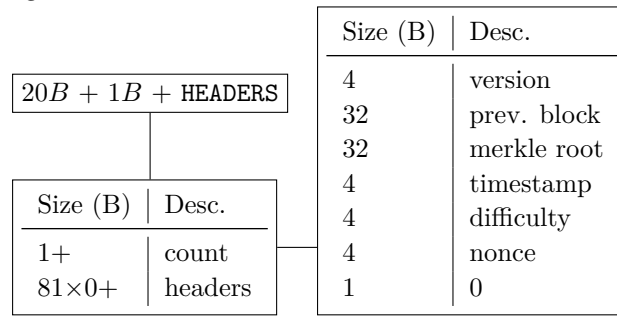

Like every packet, it contains $20$ bytes of mandatory fields, an additional
byte for the flags and the application message in the contents.
The message contains a vector of inventory objects and a variable-length field
encoding the number of objects in the vector (cf. \cref{fig:inv} for a visual
description).
The number of objects in the vector is limited to $50k$ entries which means that
the length field is always encoded in either one or three bytes.
Each inventory object is $36$ bytes big and, therefore, the smallest non-empty
\code{INV} message is $37$ bytes big.
Therefore, a \gls{TCP} payload containing a non-empty encrypted \code{INV}
message has a minimum length of $58$ bytes.
Similarly, the maximum length of such an unfragmented \gls{TCP} payload is just
slightly larger than $1.8$ megabytes.

To that end, we can use the number of bytes in the \gls{TCP} payload to
determine if a payload contains an \code{INV} message.
Given a \gls{TCP} payload with $x$ bytes, the payload can be determined to be an
\code{INV} message if $x$, after subtracting $21$ bytes for the
mandatory fields and either $1$ byte or $3$ bytes for the length field, is a
multiple of $36$, \ie if $(x-21-\{1,3\}) \bmod 36 = 0$.
This formula also gives us the number of \code{INV\_VEC} objects in the payload.
We remark that the classification accuracy of this approach is limited in
cases where message formats are identical.
Such message types, \eg \code{GETDATA} and \code{INV}, only differ in the value
of the type field and are therefore indistinguishable based on just the payload
size.

We therefore recorded the type and size of all \gls{V2} messages that were
either sent or received by the node to examine if and to what extent
\gls{V2} protects against such attacks.
We recorded a total of \measurementsnuminmsgs incoming and
\measurementsnumoutmsgs outgoing V2 messages during the measurement period.
\cref{tab:msg_sizes}
shows the three most common sizes and their respective occurrence rates as a
percentage of all messages of the specific type for select message types.
Note that the sizes refer to the entire \gls{TCP} payload, \ie the size includes
the length descriptor, flags, type and tag fields.
A general observation we make is that messages of the same size tend to occur
frequently, which is not necessarily a given since a majority of application
messages contain variable-length fields.
The \code{BLOCK} message type appears to be an exception,
with a maximum occurrence rate of $2.09\%$ at a payload size of $237$~bytes.

As a follow up to the hypothesis in
\cref{subsec:classification}, we examine the
recorded lengths of \code{INV} messages.
Notably, $14.48\%$ of the \code{INV} messages were transported in $58$-byte
payloads.
These messages contain one \code{INV\_VEC} and are the smallest possible
\code{INV} messages.
The second-most frequent payload size for \code{INV} messages was $94$ bytes.
As described in \cref{subsec:classification}, we can deduce that such a payload
contains two \code{INV\_VEC} objects.
We can apply similar logic to the other payload sizes and try to find out what
type of message is being transmitted.
For example, a $103$-byte \gls{TCP} payload could be a \code{HEADERS} message
containing exactly one block header which, as evident in \cref{tab:msg_sizes}, is
transmitted very frequently.
We refer the reader to \cref{fig:proto_msgs} for an overview of how
exactly these messages are defined.
While we cannot decipher the exact contents of the payload, previous works have
shown that such side channels can be exploited for effective attacks,
\eg\cite{ndolo2024payment, arx2023revelio, wang2025offpath}.
We believe that knowledge of the message type alone is sufficient for censorship
or delay attacks.

\begin{table}
    \caption{The three most frequent sizes of common Bitcoin \gls{V2} messages.
        The length of a \gls{TCP} payload containing such a message is equal to
        the listed length.
    }
    \label{tab:msg_sizes}
    \begin{minipage}[t]{.3\linewidth}
        \small
        \vspace{0pt}
        \begin{filecontents*}{./data/packet_sizes.csv}
type,size,count,percent
addrv2,37,1680683,34.06
addrv2,65,492838,9.99
addrv2,52,404759,8.2
block,237,5386,2.09
block,236,2118,0.82
block,287,173,0.07
blocktxn,230,8000,11.48
blocktxn,568,1321,1.9
blocktxn,530,1094,1.57
cmpctblock,23216,155,0.08
getblocktxn,55,12222,19.52
getblocktxn,57,7591,12.13
getblocktxn,56,7063,11.28
getdata,58,8090246,43.77
getdata,94,3484150,18.85
getdata,130,2001451,10.83
headers,103,330549,95.55
headers,162024,7308,2.11
headers,184,2648,0.77
inv,58,14642157,14.48
inv,94,11734755,11.61
inv,130,10061962,9.95
ping,29,4122704,100.0
pong,29,4111559,100.0
tx,375,6106782,9.52
tx,243,5386409,8.4
tx,244,4237263,6.61
version,135,65174,90.58
version,150,5319,7.39
version,136,966,1.34
\end{filecontents*}
\csvreader[
            tabular={lrr},
            table head=\toprule \textbf{Type} & \textbf{Size (B)} & \textbf{\%} \\\midrule,
            table foot=\bottomrule,
            filter={\value{csvrow}<10},
        ]
        {./data/packet_sizes.csv}{1=\type,2=\size,4=\percent}
        {\code{\expandafter\MakeUppercase\expandafter{\type}} & \size & \percent}
    \end{minipage}%
    \hfill
    \begin{minipage}[t]{.3\linewidth}
        \small
        \vspace{0pt}
        \csvreader[
            tabular={lrr},
            table head=\toprule \textbf{Type} & \textbf{Size (B)} & \textbf{\%} \\\midrule,
            table foot=\bottomrule,
            filter expr={
                test{\ifnumgreater{\thecsvinputline}{11}}
                    and
                test{\ifnumless{\thecsvinputline}{22}}
            },
        ]
        {./data/packet_sizes.csv}{1=\type,2=\size,4=\percent}
        {\code{\expandafter\MakeUppercase\expandafter{\type}} & \size & \percent}
    \end{minipage}%
    \hfill
    \begin{minipage}[t]{.3\linewidth}
        \small
        \vspace{0pt}
        \csvreader[
            tabular={lrr},
            table head=\toprule \textbf{Type} & \textbf{Size (B)} & \textbf{\%} \\\midrule,
            table foot=\bottomrule,
            filter expr={
                test{\ifnumgreater{\thecsvinputline}{21}}
            },
        ]
        {./data/packet_sizes.csv}{1=\type,2=\size,4=\percent}
        {\code{\expandafter\MakeUppercase\expandafter{\type}} & \size & \percent}
    \end{minipage}
\end{table}

In practice, classification at the network level is not always as
straightforward and accurate for the following reasons:
\begin{enumerate}
    \item application messages naturally add some randomness to the bytestream
        because of the presence of variable-length fields in most message types.
        As related works have shown, message types with distinct but invariable
        lengths can be classified extremely accurately despite
        encryption~\cite{ndolo2024payment, arx2023revelio}; and
    \item application messages are transmitted as continuous streams of data that
        are separated by magic bytes, and therefore, a \gls{TCP} payload may
        carry more than one application message. Unlike V1 streams, packet
        boundaries cannot be determined by examining the payload.
\end{enumerate}
However, we contend that the frequencies are sufficient for network-level
classification attacks using just the length of \gls{TCP} payload.

\section{Implementation: Testbed, attacks \& simulator}
\label[appendix]{sec:implementation}

In the following, we describe the testbed that
we used for our analysis of the attacks (\cref{subsec:testbed}).
We also provide further details on our implementation
of the eclipse attack
(\cref{subsec:impl_eclipse}), address table
simulator used to evaluate the duration of the eclipse attack
(\cref{subsec:impl_addrsim})
and downgrade attack
(\cref{subsec:impl_downgrade}).

\subsection{Testbed}
\label[appendix]{subsec:testbed}

Accurately testing network-layer attacks at scale is challenging for several
reasons.
It requires that node behaviour resembles real implementations as well as a
realistic representation of the networking layer.
Furthermore, and especially in the case of partitioning attacks, the size of the
network must be taken into account.
This requires resources that are not always available.
Related attacks have in the past been evaluated by either simulating certain
aspects of the Bitcoin protocol~\cite{tran2020stealthier}, attacking nodes
deployed in the public network~\cite{fan2021conman, heilman2015eclipse} or
setting up a testbed of real Bitcoin Core clients running in a dedicated
network~\cite{apostolaki2017hijacking}.
We decided to perform our evaluations in a private network as it provides the
most authentic client behaviour without the risk of interfering with third-party
nodes in any way.
Unfortunately, some of the components of the testbed used
in~\cite{apostolaki2017hijacking} are not publicly available.
While \code{warnet},\footnote{\url{https://github.com/bitcoin-dev-project/warnet}}
a software tool to generate private Bitcoin networks, is promising, it does not
provide access to key networking functionality, \eg defining
different IP networks.

We therefore built our own testbed, named \testbed, out of necessity, and make
it publicly available.\footnote{\testbedurl}
It uses Containernet~\cite{peuster2016medicine}, a fork of the Mininet
project~\cite{lantz2010network}, to build a virtual network topology that
runs on a Linux machine.
Containernet extends Mininet with support for Docker containers as hosts.
The testbed has the potential to support a wide range of network-layer research on
Bitcoin as well as related networks and is highly configurable.
Like \code{warnet}, Docker containers running configurable versions of Bitcoin
Core in \enquote{regtest} mode are used as hosts and are connected to form a
private Bitcoin network.
It handles bootstrapping automatically using three \btcversion nodes serving as
seed nodes.
It includes a Prometheus server and Grafana dashboard monitoring various
application-specific metrics, \eg peer count, as well as relevant system metrics
of the Linux server.
It supports building a network topology of multiple \glsentryplural{AS} that are
connected via switches and routers.
Currently, different networks are reserved for various types of hosts:
$175.0.0.0/8$ for regular Bitcoin core clients, $193.168.1.0/24$ for a victim
node as well as seeds and $11.0.0.0/8$ for adversarial clients.
Parameters such as the number of subnets, nodes per subnet and block generation
rate are all configurable at runtime.
This ensure scalability and general accessibility to users with different
computation resources to their disposal.

We plan on extending \testbed and adding support for more features in the
future.
We intend on modelling the network topology to resemble the public Bitcoin
network more, \eg the distribution of \gls{AS} and latencies between peers.
We also plan on generalising \testbed such that it can be used for various
\gls{P2P} networks aside from Bitcoin.

\subsection{Eclipse attack}
\label[appendix]{subsec:impl_eclipse}

Our implementation of the eclipse attack consists of two components: the
network-level attack logic and the application-level logic that handles the
adversary's connections with the victim.

\subsubsection{Attack}

We implemented a prototype of the attack using the
\netfilter\footnote{\url{https://www.netfilter.org}} framework which has been
used in the past by related works~\cite{apostolaki2017hijacking,
ndolo2024payment}.
The attack code was implemented in user space using the
\code{nfq-rs}\footnote{\url{https://github.com/nbdd0121/nfq-rs}} library in
Rust.
The source code is available for reviewers in an online repository.%

The program begins by attaching itself to a queue; traffic to/from the victim
node should be directed to this queue.
The program monitors all the traffic and issues an \code{ACCEPT} verdict in
almost all cases.
The only exception is for \gls{TCP} \syn{}s subsequent to a replayed payload.
It stores the first payload that is suspected to be a \code{PING} from the
victim in a map in which the destination is the key.
When a second packet transporting a \code{PING} is identified, its payload is
replaced with the stored payload, the \gls{TCP} checksum is recalculated and an
\code{ACCEPT} verdict is issued.
The payload is deleted afterwards such that it only holds a maximum of one
$29$-byte payload per connection at any given point.

\subsubsection{Bitcoin \gls{P2P} client}

The adversary requires a lightweight Bitcoin client that is able to establish
and maintain \gls{P2P} connections with the victim node.
As we did not find any such client, we implemented our own client in Rust using
the
\code{rust-bitcoin}\footnote{\url{https://github.com/rust-bitcoin/rust-bitcoin}} library.
The result is a \emph{minimal} Bitcoin client, named \btcclient~that implements
the aspects of the \gls{P2P} transport protocol described in
\cref{sub:sustaining_attack}.
The code is publicly available in an online repository.\footnote{\pocurl}
\btcclient can be used in client mode, \ie initiates a connection to a node, or
in server mode, \ie a \gls{TCP} server listens for and accepts connections.
Due to the backwards compatibility specified in \relevantbip, the client does
not need to support \gls{V2}.

\subsection{Simulation of address selection}
\label[appendix]{subsec:impl_addrsim}

In order to evaluate how long the eclipse attack would require in practice, we
implemented a simulator for the selection of outgoing addresses.
It was implemented in Rust and is available in a repository.\footnote{\simurl}
The simulator expects a dump of a Bitcoin Core node's address table, the number
of attacker addresses, the number of connections to close per time interval and
a seed for the random number generator as input. It generates the specified
number of attacker addresses in the $0.0.0.0/8$ address space which we chose in
order to not produce addresses that may collide with addresses in the provided
database.
The simulation begins by selecting ten addresses at random as the initial peers
and continues to keep track of current peers throughout the simulation.
The address database is then populated with the attacker addresses in
random buckets of the new table.
Following the Bitcoin protocol, the simulator makes a feeler connection every two
minutes to a random address and moves the address to the tried table.
In case the address was present multiple times, all references are also deleted
from the new table.
Every four minutes, the specified number of random, non-adversarial connections
are closed.
This means that the simulator fills the freed slots with equally many new
connections.
It does so by first selecting either the new or tried table at random, and then
selecting a random address from the chosen table.
The address, and all of its references, are removed from the table.
The simulation runs until all ten connections belong to the generated addresses.

\subsection{Downgrade attack}
\label[appendix]{subsec:impl_downgrade}

We implemented a prototype of the downgrade attack using the
\netfilter framework as a user-space program in Rust.
The source code is available for reviewers in an online repository.%
The basics of the implementation are very similar to the implementation of the
eclipse attack described in Appendix~\ref{subsec:impl_eclipse}.

The program begins by attaching itself to a queue; traffic to/from the victim
node are directed to this queue.
It monitors all the traffic it receives and issues an \code{ACCEPT} verdict for
all packets.
When a \gls{TCP} \syn~in either direction is taken from the queue, the program
starts to keep track of the particular connection.
Specifically, it waits for the \gls{TCP} handshake to be completed and then
monitors the payloads being transmitted over the connection.
If the first packet's payload contains more than $126$ bytes, the program alters
the second packet to a \gls{TCP} \code{RST} packet.
It sets the \code{RST} flag and recalculates the \gls{TCP} checksum before
returning the packet to the kernel.

\section{Supplementary material}

The following is a follow-up of \cref{subsec:outgoing}; we expound on what the attacker
needs to do in order to occupy the victim node's outgoing connection slots.

\label[appendix]{subsec:outgoing_connections}

\subparagraph{Entering the tables}

The most fundamental step the adversary must perform is to insert their
addresses into the victim's new table.
Whether or not or when such an address moves to the tried table is beyond the
adversary's control.
Our adversary inserts its malicious addresses by participating in the \gls{P2P}
network and relaying its addresses via regular \code{ADDR} to the victim node.
The most effective strategy is to propagate these addresses to the victim
directly as well as to other nodes in the network.
In the latter case, the addresses will eventually reach the victim via gossip.

\subparagraph{Prefix groups}

The Bitcoin protocol stipulates that each of a node's outgoing connections must
be from a different prefix group.
The adversary must therefore control at least \emph{ten} IP addresses from
distinct subnets that are waiting for connections.
Feeler connections do not partake in information relay in any way, but trigger
the migration of addresses to the smaller tried table.
Therefore, it is in the adversary's best interest to provide more than the bare
minimum number of addresses in case they are moved to the tried table.
Hence, the more unique prefix groups the adversary controls, the faster they
will be able to eclipse a node.
As our threat model assumes a network-level adversary such as an \gls{AS}, we do
not consider this requirement to be inherently deterring.

\subparagraph{Table size}

As the new table can hold up to $2^{16}$ entries, the odds of an adversarial
address being selected are not high.
Our adversary improves their chances by ensuring that each address occurs
multiple times in the new table and, most importantly, that each of their
addresses is from a different prefix group.
While our attack relies more on the target depleting the entries rather than
filling the entire table, we refer to~\cite{tran2020stealthier} for a related
discussion.

\subparagraph{Desirable services}

Once an address has been selected as a potential outgoing connection, Bitcoin
Core performs some
tests\footnote{\url{https://github.com/bitcoin/bitcoin/blob/28.x/src/net.cpp\#L2452-L2791}}
on the address before initiating a connection request.
Among others, it verifies that the address is valid, publicly routable and
belongs to a network group that is distinct from the current outbound network
groups.
The logic also verifies that the potential peer provides all of the required
application-level services.
Services are features that a node supports, \eg \code{NODE\_P2P\_V2} for
\gls{V2},
and are advertised in \code{ADDR} messages.
All the adversary needs to do is to advertise the relevant flags in the
\code{ADDR} messages they use to relay addresses.
They do not have to actually offer any of the announced services as there is
no verification or validation mechanism.
They should announce the \code{NODE\_NETWORK},
\code{NODE\_NETWORK\_LIMITED} and \code{NODE\_WITNESS} flags to cover all
relevant cases.

\begin{figure}[h]
    \centering
    \begin{subfigure}[t]{0.475\textwidth}
        \includegraphics[height=4.5cm]{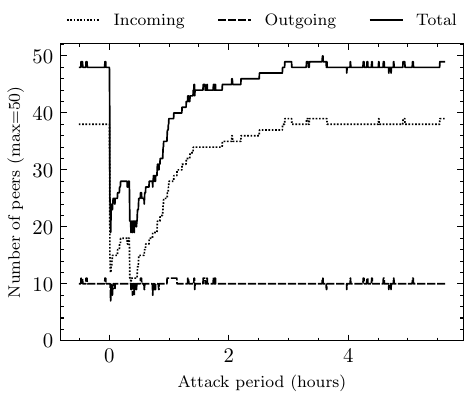}
        \caption{
            The number of peers the victim node had before and during the attack.
        All but one of its incoming connection slots were occupied at the start of the
        attack.
        The initial decline in incoming connections is expected because the
        node already had multiple connections that the attacker tampered with.
        The results show that the attack progresses quite fast initially, but needs more
        time to fill the last slots.
        The victim node was eclipsed after five hours in this experiment.
        }
        \label{fig:target_peer_count}
    \end{subfigure}
    \hfill
    \begin{subfigure}[t]{0.475\textwidth}
        \includegraphics[height=4.5cm]{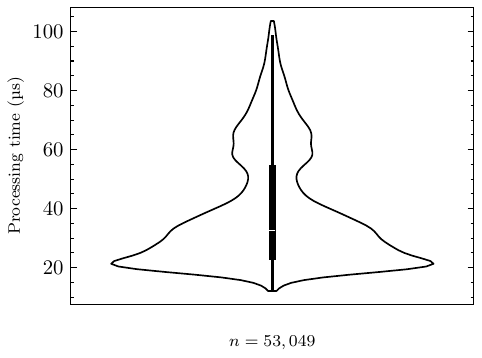}
        \label{fig:delay_eclipse}
        \caption{
            The packet processing times for the prototype
        implementation of the eclipse attack.
        The data was collected during the first five hours of the attack, and is
        depicted as a violin plot.
        This is the duration from the program accessing a packet from the queue to it
        returning the (altered) packet to the kernel.
        The implementation is quite efficient and issues a verdict on a packet in a
        median time of $\approx\micros{32}$ and mean time of $\micros{39}$.
        }
    \end{subfigure}
    \caption[]{Additional results from the evaluation of the eclipse attack.}
    \label{fig:supp_eclipse}
\end{figure}%

\begin{figure}[h!]
    \centering
    \begin{subfigure}[t]{0.475\textwidth}
    \includegraphics[height=4.5cm]{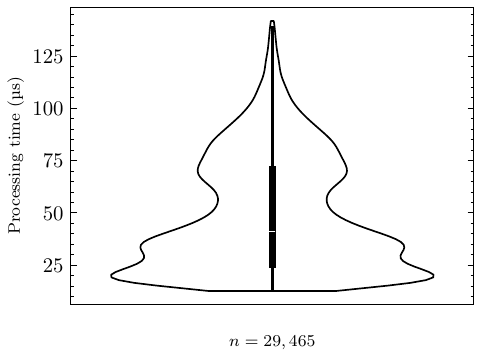}
    \caption{
        The packet processing times for the prototype
        implementation of the downgrade attack based on packets received during the
        first two hours of the attack.
        The figure shows the time from the program accessing a packet from
        the queue to it returning the (altered) packet to the kernel.
        The implementation is relatively efficient and issues a verdict on a
        packet within a median time of $\approx\micros{41}$ and mean time of
        $\micros{49}$.
    }
    \label{fig:delay_downgrade}
    \end{subfigure}
    \hfill
    \begin{subfigure}[t]{0.475\textwidth}
    \includegraphics[height=4.5cm]{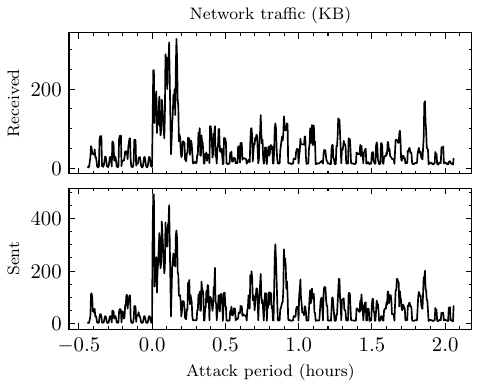}
    \caption{
        The amount of data sent and received by the victim's Bitcoin Core client
        before and during the attack in kilobytes.
        The figure shows an unmistakable peak in both sent and received traffic when the
        attack starts.
        This is because many connections are closed and opened in quick succession at
        that point due to the restart.
    }
    \label{fig:target_traffic_downgrade}
    \end{subfigure}
    \caption[]{Additional results from the evaluation of the downgrade attack.}
    \label{fig:supp_downgrade}
\end{figure}

\begin{table}
    \caption[]{Summary of \relevantbip's objectives and how they are achieved.
    Traffic shaping is considered to be out-of-scope of \relevantbip and is
    not implemented. Upgradability is not yet relevant as there is currently
    only one version of \gls{V2}.}
    \label{tab:v2_goals}
    \centering\small
    \begin{tabular}{lp{.145\linewidth}p{.37\linewidth}p{.37\linewidth}}
        \toprule
        & \textbf{Objective} & \textbf{What} & \textbf{How} \\
        \midrule
        \#1 & Confidentiality against passive attacks & An attacker must not be
        able to determine the plaintext being exchanged from a recorded \gls{V2}
        bytestream without additional information. & Unauthenticated
        ChaCha20Poly1305 encryption of the plaintext communication between
        nodes. \\
        \#2 & Observability of active attacks & An active man-in-the-middle
        attacker runs the risk of being detected as a session ID uniquely
        identifies an encrypted channel. & Manual comparison of session IDs by
        operators. \\
        \#3 & Pseudorandom bytestream & A passive attacker must not be able to
        distinguish a bytestream without timing and fragmentation information
        from a uniformly random bytestream. & ChaCha20 encryption of
        the length field with a different (from plaintext encryption) key. \\
        \#4 & Shapable bytestream & It should be possible to shape \gls{V2}
        bytestreams to increase resistance to traffic analysis. & The ignore bit
        in encrypted packets is used to indicate decoy packets. \relevantbip
        does not specify how and when decoy packets should be used. \\
        \#5 & Forward secrecy & An eavesdropper who compromises a node's sessions
        secrets should not be able to decrypt historical session traffic. & A
        hash step is performed every $224$ messages ($\leq~3.11$ days) to rekey
        the encryption ciphers. \\
        \#6 & Upgradability & The protocol should provide an upgrade path using
        transport versioning which can be used to add features in the future. &
        The \code{VERSION} message sent during the version negotiation phase is
        used to encode support for a specific protocol version. \relevantbip
        does not specify how exactly version support should be announced. \\
        \#7 & Compatibility & \gls{V2} clients accept inbound V1 connections to
        minimise risk of network partitions. & Connection attempts are to be
        retried in case of immediate disconnection when establishing a \gls{V2}
        connection. \\
        \#8 & Low overhead & The computational cost or bandwidth for nodes that
        implement \gls{V2} should not increase substantially compared to the
        original protocol & The protocol provides a slight bandwidth reduction
        compared to the original protocol. Among other measures, the format
        of \gls{V2} messages leads to smaller messages. \\
        \bottomrule
    \end{tabular}
\end{table}

\end{document}